\documentclass{LMCS}

\usepackage{diagrams}
\diagramstyle[size=6mm]
\newarrow{Into} {boldhook}--->
\newarrow{Line} -----
\newarrow{Myto} ---->
\newarrow{Dotsto} ....>
\newarrow{Equals} =====
\theoremstyle{plain}
\newtheorem{lemdefn}[thm]{Lemma and Definition}

\usepackage{amsmath}
\usepackage{amssymb}
\usepackage{xspace}
\newcommand{\CASL}{\textsc{Casl}\xspace}
\newcommand{\CoCASL}{\textsc{CoCasl}\xspace}

\newlength{\croutw}
\newlength{\crouth}
\newcommand{\crossout}[1]%
        {\settowidth{\croutw}{$#1$}\settoheight{\crouth}{$#1$}#1%
        \hspace{-1.0\croutw}\raisebox{0.3\crouth}{\rule{\croutw}{0.1ex}}}

\newcommand{\commentout}[1]{\ignorespaces}

\newcommand{\Pow}{\mathcal{P}}

\newcommand{\bit}{\begin{itemize}}
\newcommand{\eit}{\end{itemize}}

        {\begin{enumerate}}%
        {\end{enumerate}}
        {\begin{enumerate}}%
        {\end{enumerate}}
        {\begin{enumerate}}%
        {\end{enumerate}}
        {\begin{enumerate}}%
        {\end{enumerate}}
\newcommand{\Cat}{\mathbf}
\newcommand{\Cls}{\mathcal}

\newcommand{\BC}{{\Cat C}}

\newcommand{\CM}{{\Cls M}}

\newcommand{\CT}{{\Cls T}}

\newcommand{\into}{\hookrightarrow}
\newcommand{\impl}{\Rightarrow}

\newcommand{\argument}{\_\!\_}%{\underline{\;\;}}
\newcommand{\Set}{\Cat{Set}}

\newcommand{\HasCASL}{{\sc HasCasl}\xspace}

\newcommand{\sbullet}{\mbox{ $\scriptstyle\bullet$ }}
\newcommand{\lambdaTotal}[2]{\lambda\, #1\sbullet\!\! !\, #2}

\newcommand{\UnitType}{{\mathit{Unit}}}

\newcommand{\Logical}{{\mathit{Logical}}}

\newcommand{\TypeKind}{\mathit{Type}}

\newcommand{\exeq}{\stackrel{e}{=}}
\newcommand{\IsDef}{\operatorname{def}}

\providecommand{\pfun}{\mathrel{\to?}}

\newcommand{\cArrow}{\stackrel{c}{\longrightarrow}}
\newcommand{\pcArrow}{\cArrow ?}

        {\smallskip\par\noindent\qquad$\begin{array}{@{}l@{{}={}}l}}%
        {\end{array}$\par\noindent}

\newlength{\myboxwidth}
\newenvironment{myfigure}{\begin{figure}\begin{center}
	\setlength{\fboxsep}{10pt}}%
        {\end{center}\end{figure}}

\newcommand{\Nat}{{\mathbb{N}}}

\newcommand{\secHasCASL}{H{\small AS}C{\small ASL}\xspace}

\newcommand{\tfun}{\to}

\newcommand{\Type}{\textbf{type}\xspace}
\newcommand{\Free}{\textbf{free}\xspace}
\newcommand{\Generated}{\textbf{generated}\xspace}
\newcommand{\mi}[1]{\mathit{#1}}
\newcommand{\Group}	{\textbf{\{}}
\newcommand{\EndGroup}	{\textbf{\}}}
\newcommand{\lambdaTerm}[2]{{\lambda #1\sbullet\,#2}}
\newcommand{\resOp}{\restriction}

\newcommand{\ReRe}{\Cat{ReRe}}

\newcommand{\parr}{\rightharpoonup}
\newcommand{\pmCat}{\Cat{P}}
\newcommand{\Land}{\bigwedge}
\newcommand{\Th}{\mathsf{Th}}
\newcommand{\Cl}{\mathsf{Cl}}
\newcommand{\obsep}{.\,}
\newcommand{\Sg}{\mathit{Sg}}
\newcommand{\smalliff}{\Leftrightarrow}
\newcommand{\infle}{\sqsubseteq}
\newcommand{\cposup}{\bigsqcup}

\newcommand{\Pol}{P}

\newcommand{\CA}{\mathcal{A}}
\newcommand{\CB}{\mathcal{B}}

\newcommand\dash{\nobreakdash-\hspace{0pt}}

\usepackage{hetcasl}
\usepackage{enumerate,hyperref}

\newcommand{\eat}[1]{}
\def\doi{4 (4:17) 2008}
\lmcsheading%
{\doi}
{1--27}
{}
{}
{Dec.~15, 2007}
{Dec.~25, 2008}
{}   

\begin{document}

\title[Bootstrapping Inductive and Coinductive Types in HasCASL]{Bootstrapping Inductive and Coinductive Types\\in HasCASL}
\author[L.~Schr{\"o}der]{Lutz Schr{\"o}der}
\address{DFKI Bremen and Department of Computer Science, 
  University of Bremen, Germany}
\email{Lutz.Schroeder@dfki.de}
\thanks{Work performed under DFG-project
    \HasCASL (KR 1191/7-2) and BMBF-project FormalSafe (FKZ 01IW07002)}



\titlecomment{This work is an extended version of~\cite{Schroder07a}}

\hyphenation{quasi-top-os quasi-top-oses}
\begin{abstract}
  \noindent We discuss the treatment of initial datatypes and final
  process types in the wide-spectrum language HasCASL. In particular,
  we present specifications that illustrate how datatypes and process
  types arise as bootstrapped concepts using HasCASL's type class
  mechanism, and we describe constructions of types of finite and
  infinite trees that establish the conservativity of datatype and
  process type declarations adhering to certain reasonable
  formats. The latter amounts to modifying known constructions from
  HOL to avoid unique choice; in categorical terminology, this means
  that we establish that quasitoposes with an internal natural numbers
  object support initial algebras and final coalgebras for a range of
  polynomial functors, thereby partially generalising corresponding
  results from topos theory. Moreover, we present similar
  constructions in categories of internal complete partial orders in
  quasitoposes.
\end{abstract}

\maketitle             

\section*{Introduction}
\noindent The formally stringent development of software in a unified
process calls for wide-spectrum languages that support all stages of
the formal development process, including abstract requirements,
design, and implementation. In the \CASL language
family~\cite{CASL-UM}, this role is played by the higher-order \CASL
extension
\HasCASL~\cite{SchroderMossakowski02,SchroderMossakowski08HCOverview}. Like
in first-order \CASL, a key feature of \HasCASL is support for
inductive datatypes, which appear in the specification of the
functional correctness of software. In the algebraic-coalgebraic
language \CoCASL~\cite{MossakowskiEA04}, this concept is complemented
by coinductive types, which appear as state spaces of reactive
processes. Many issues revolving around types of either kind gain in
complexity in the context of the enriched language \HasCASL; this is
related both to the presence of additional language features such as
higher order types and type class polymorphism and to the nature of
the underlying logic of \HasCASL, an intuitionistic higher order logic
of partial functions without unique choice which may, with a certain
margin of error, be thought of as the internal logic of quasitoposes
(more precisely, it is the internal logic of partial cartesian closed
categories with equality~\cite{Schroder04,Schroder06a}).

Here, we discuss several aspects of \HasCASL's concept of inductive
datatype, as well as the perspective of adding coinductive types to
\HasCASL. To begin, we present the syntax and semantics of inductive
datatypes, which may be equipped with reachability constraints or
initiality constraints; both types of constraints may be relatively
involved due to the fact that constructor arguments may have complex
composite types. We then go on to show how initial datatypes may be
specified in terms of \HasCASL's type class mechanism. On the one hand,
this shows that initial datatypes need not be regarded as a built-in
language feature, but may be considered as belonging into a `\HasCASL
prelude'. On the other hand, the specifications in question give a good
illustration of how far the type class mechanism may be stretched. We
then briefly discuss how a simple dualisation of these specifications
describes final process types in the style of \CoCASL; thus, the
introduction of such types into \HasCASL would merely constitute
additional syntactic sugar (although concerning the relationship to
\CASL and \CoCASL, for both datatypes and process types certain caveats
apply related to \HasCASL's Henkin semantics).

Finally, we tackle the issue of the conservativity of datatype and
process type declarations. We follow the method employed in standard
HOL~\cite{Paulson97,BerghoferWenzel99}, which consists in defining a
universal type of trees and then carving out the desired inductive or
coinductive types. However, the constructions need to be carefully
adapted in order to cope with the lack of unique choice. Abstracting
our results to the categorical level, we prove, in partial
generalisation of corresponding results for
toposes~\cite{PareSchumacher78,MoerdijkPalmgren00}, that any
quasitopos (indeed, any partial cartesian closed category with
equality and finite coproducts) with nno supports initial algebras and
final coalgebras for certain classes of polynomial functors. Moreover,
we obtain corresponding results for datatypes and process types
equipped with complete partial orders (the former are called free
domains in \HasCASL). These types serve as the correspondent of
programming language datatypes in \HasCASL's internal modelling of
denotational semantics.

The material is organised as follows. We recall some aspects of the
syntax and semantics of \HasCASL, including the relationship between
\HasCASL' Henkin models and categorical models, in
Sect.~\ref{sec:hascasl}. In Sect.~\ref{sec:datatypes}, we expand on
the semantics of generated and free datatypes in \HasCASL. These two
sections summarise material
from~\cite{Schroder06a,SchroderMossakowski08HCOverview}. We then go on
to present the bootstrapped specification of the syntax and semantics
of signature functors and inductive datatypes using the type class
mechanism in Sect.~\ref{sec:bootstrap}. In Sect.~\ref{sec:cotypes} we
discuss how these concepts extend naturally to coinductive process
types. We present the constructions establishing the conservativity of
datatype and process type declarations in
Sect.~\ref{sec:cons}. Finally, we recall the modelling of general
recursive programs by means of an adapted version of domain theory in
Sect.~\ref{sec:domains}, and show how the constructions of plain
initial datatypes and final process types can be modified to obtain
corresponding constructions on domains.

\section{\secHasCASL}\label{sec:hascasl}

\noindent The wide-spectrum language
\HasCASL~\cite{SchroderMossakowski02} extends the standard algebraic
specification language \CASL by intuitionistic partial higher order
logic, equipped with a set-theoretic Henkin semantics, an extensive
type class mechanism, and HOLCF-style support for recursive
programming. \HasCASL moreover provides support for
functional-imperative specification and programming in the shape of
monad-based computational
logics~\cite{SchroderMossakowski03a,SchroderMossakowski04,SchroderMossakowski04b,WalterEA05}.
Tool support for \HasCASL is provided in the framework of the Bremen
heterogeneous tool set Hets~\cite{MossakowskiEA07}. We expect the
reader to be familiar with the basic \CASL syntax (whose use in our
examples is, at any rate, largely self-explanatory), referring
to~\cite{CASL-UM,CASL-RM} for a detailed language description. Below,
we review the \HasCASL language features most relevant for the
understanding of the present work, namely type class polymorphism and
certain details of \HasCASL's higher order logic;
see~\cite{SchroderMossakowski08HCOverview} for a full language
definition. Moreover, we recall \HasCASL's Henkin semantics and its
relation to categorical models in quasitoposes and, more generally,
partial cartesian closed categories with equality~\cite{Schroder06a}.

\subsection{The internal logic of \HasCASL}
\label{sec:intlog}
The logic of \HasCASL is based on the partial $\lambda$\dash
calculus~\cite{Moggi86}. It is distinguished from standard HOL by
having intuitionistic truth values and partial function types $t\pfun
s$ (besides total function types $t\tfun s$); $\lambda$\dash
abstractions $\lambdaTerm{x:t}{\alpha}$ denote partial functions,
i.e.\ inhabitants of partial function types $t\pfun s$, while total
$\lambda$\dash abstractions, inhabiting the total function type
$t\tfun s$, are denoted $\lambdaTotal{x:s}{\alpha}$. There are
moreover a unit type $\UnitType$, with unique inhabitant $()$, and
product types $s\times t$. Predicates then arise as partial functions
into $\UnitType$, where definedness is understood as satisfaction, and
the type of truth values is $\Logical=\UnitType\pfun\UnitType$. We
denote application of a function $f$ to an argument $x$ as $f\ x$,
under the convention that application is left-associative. As
in~\cite{Schroder06a}, we moreover denote by $\alpha\resOp\phi$ the
\emph{restriction} of a term $\alpha$ to a formula $\phi$, i.e.\
$\alpha\resOp\phi$ is defined iff $\alpha$ is defined and $\phi$
holds, and in this case is equal to $\alpha$ (essentially, $\resOp$ is
just the first projection).

In a partial setting, there are numerous readings of equality; here,
we require \emph{strong equality}, denoted $=$ and read `one side is
defined iff the other is, and in this case, both sides are equal', as
well as \emph{existential equality}, denoted $\exeq$ and read `both
sides are defined and equal'. Equality of terms in the partial
$\lambda$\dash calculus is axiomatised largely as
expected~\cite{Moggi86,Schroder06a}, with a few subtleties attached to
partiality --- e.g.\ $\beta$\dash equality
$(\lambdaTerm{x:t}{\alpha})\gamma=\alpha[\gamma/x]$ holds only if the
term $\gamma$ is defined. We assume that equality is \emph{internal},
i.e.\ that there exists on every type a binary predicate representing
existential equality, also written $\exeq$; this defines the
\emph{partial $\lambda$\dash calculus with equality}. In the partial
$\lambda$\dash calculus with equality, an intuitionistic predicate
logic is defined in the standard way (see
e.g.~\cite{Schroder04,Schroder06a}) by abbreviations such as
\begin{equation*}
  \forall x:t\sbullet \phi = ((\lambdaTerm{x:t}{\phi})\exeq
  (\lambdaTerm{x:a}{()})),
\end{equation*}
where $()$ is the unique inhabitant of $\UnitType$. The arising logic
includes higher-order universal and existential quantifiers $\forall$,
$\exists$, propositional connectives $\land$, $\lor$, $\impl$,
$\smalliff$, $\neg$, and truth values $\top$, $\bot$.

The difference between the \HasCASL logic and the more familiar topos
logic~\cite{LambekScott86} is the absence of unique
choice~\cite{Schroder06a}, where we say that a type $a$ admits
\emph{unique choice} if $a$ supports \emph{unique description} terms
of the form $(\iota x:a.\,\phi):a$ designating the unique element $x$
of $a$ satisfying the formula $\phi$ (which may of course mention
$x$), if such an element exists uniquely (this is like Isabelle/HOL's
\texttt{THE}~\cite{NipkowEA02}). In \HasCASL, the unique choice
principle may be imposed if desired by means of a polymorphic
axiom~\cite{SchroderMossakowski08HCOverview}. The lack of unique
choice requires additional effort in the construction of tree types
establishing the conservativity of datatype and process type
declarations; this is the main theme of Sect.~\ref{sec:cons}. The
motivation justifying this effort is twofold:
\begin{enumerate}[$\bullet$]
\item Making do without unique choice essentially amounts to admitting
  models in quasitoposes rather than just in toposes (see
  Section~\ref{sec:semantics}). Interesting set-based quasitoposes
  include pseudotopological spaces and reflexive relations; further
  typical examples are categories of extensional presheaves, including
  e.g.\ the category of reflexive logical relations, and categories of
  assemblies, both appearing in the context of realisability
  models~\cite{Phoa92,RosoliniStreicher99}. In particular, the
  category of $\omega$\dash sets is a quasitopos but not a topos; it is
  embedded as a full subcategory into the effective topos, whose
  objects however have a much more involved description than
  $\omega$\dash sets~\cite{Phoa92}. Quasitoposes also play a role in the
  semantics of parametric polymorphism~\cite{BirkedalMoegelberg05}.
\item A discipline of avoiding unique choice leads to constructions
  which may be easier to handle in machine proofs than ones containing
  unique description operators; cf.\ e.g.\ the explicit warning
  from \cite{NipkowEA02}, Sec.~5.10:\medskip

   \begin{quote}
     \emph{``Description operators can be hard to reason about. Novices should
     try to avoid them. Fortunately, descriptions are seldom required.''}
   \end{quote}\medskip

   \noindent As we shall discuss below, one occasion where the theory
   development of Isabelle/HOL does require description operators is
   the construction of datatypes, and our results shed some light on
   the question to which extend this can be avoided.
\end{enumerate}

\subsection{Type class polymorphism}\label{sec:poly}
\HasCASL's shallow polymorphism revolves around a notion of
type class. Type classes are syntactic subsets of \emph{kinds}, where
kinds are formed from \emph{classes}, including a base class
$\TypeKind$ of all types, and the type function arrow $\to$. Classes
are declared by means of the keyword $\textbf{class}$; e.g.\
\begin{equation*}
\CLASS \Id{Functor}<\TypeKind\to\TypeKind
\end{equation*}
declares a class $\Id{Functor}$ of type constructors, i.e.\ operations
taking types to types. Types are declared with associated classes (or
with default class $\TypeKind$) by means of the keyword $\textbf{type}$;
e.g.\ a type constructor $F$ of class $\mi{Functor}$ is declared by
writing
\begin{equation*}
\TYPE F:\Id{Functor}
\end{equation*}
Such declarations may be generic; e.g.\ if $\Id{Ord}$ is a class, then
we may write
\begin{align*}
&\VAR a,b:\Id{Ord}\\
&\TYPE a\times b:\Id{Ord}
\end{align*}
thus imposing that the class $\Id{Ord}$ is closed under products; note
how the keyword $\textbf{var}$ is used for both standard variables and
type variables. Operations and axioms may be polymorphic over any class,
i.e.\ types of operations and variables may contain type variables with
assigned classes.

In order to ensure the institutional satisfaction condition
(invariance of satisfaction under change of notation), polymorphism is
equipped with an \emph{extension semantics}~\cite{SchroderEA04}; the
only point to note for purposes of this work is that as a consequence,
a specification extension is, in \CASL terminology,
(model-theoretically) conservative, i.e.\ admits expansions of models,
iff it only introduces names for entities already expressible in the
present signature. In the case of types, this means that e.g.\ a
datatype declaration is conservative iff it can be implemented as a
subtype of an existing type.

\subsection{Henkin Models and Partial Cartesian Closed Categories}
\label{sec:semantics}

\noindent The set-theoretic semantics of \HasCASL is given by
\emph{intensional Henkin models}, where function types are equipped
with application operators but are neither expected to contain all
set-theoretic functions nor indeed to consist of functions; in
particular, different elements of the function type may induce the
same set-theoretic function. Such models are essentially equivalent to
models in (varying!) partial cartesian closed categories (pccc's) with
equality~\cite{Schroder06a}; these categories are slightly more
general than quasitoposes~\cite{AdamekHerrlich90}, which can be seen
as finitely cocomplete pccc's with equality. Below, we summarise some
of the details of the categorical viewpoint; we refer
to~\cite{Schroder06a,SchroderMossakowski08HCOverview} for the full
definition of intensional Henkin models. 

A \emph{dominion}~\cite{RosoliniThesis} on a category $\BC$ is a class
$\CM$ of monomorphisms in $\BC$ which contains all identities and is
closed under composition and pullback stable, the latter in the sense
that pullbacks, or inverse images, of $\CM$\dash morphisms along
arbitrary morphisms exist and are in $\CM$. The pair $(\BC,\CM)$ is
called a \emph{dominional category}. A \emph{partial morphism} $X\parr
Y$ in $(\BC,\CM)$ is a span $X\lInto^m D \rTo^f Y$, where $m\in\CM$,
taken modulo isomorphic change of $D$. Partial morphisms $(m,f)$ are
composed by pullback formation. Intuitively, $(m,f)$ is a partial map
defined on the subobject $D$ of $X$.  The partial morphisms in
$(\BC,\CM)$ form a category $\pmCat(\BC,\CM)$ (with small hom-sets if
$\BC$ is $\CM$-wellpowered~\cite{AdamekHerrlich90}), into which $\BC$
is embedded by mapping a morphism $f$ to the partial morphism
$(id,f)$. If $\BC$ is cartesian, i.e.\ has a terminal object $1$ and
binary products $A\times B$, then $\BC$ is a \emph{partial cartesian
  closed category (pccc)} if the functor
\begin{equation*}
  \BC\rTo^{\argument \times A} \BC\into \pmCat(\BC,\CM)
\end{equation*}
has a right adjoint for each object $A$ in $\BC$. If in addition,
$\CM$ contains all diagonal morphisms $A\to A\times A$, then a
monomorphism $f$ in $\BC$ is extremal iff $f$ is regular iff $f\in\CM$
(see e.g.~\cite{AdamekHerrlich90} for definitions of extremal and
regular monomorphisms). In this case, $\BC$ is called a \emph{pccc
  with equality}.
\begin{rem}
  The above constructive approach to partial maps is complemented by a
  variety of direct approaches which take the category of partial maps
  as basic and axiomatise its properties. The details of these
  approaches and their relationship to the constructive approach are
  discussed in some breadth in~\cite{Schroder06a}.
\end{rem}

\noindent It has been shown in~\cite{Schroder06a} that one has an
equivalence between theories in the partial $\lambda$\dash calculus
with equality and pccc's with equality. Here, a \emph{theory} consists
of a set of basic types, from which composite types are obtained
inductively by forming partial function types $s_1\times\dots\times
s_n\pfun t$, a set of basic operations with assigned types, and a set
of axioms, expressed as \emph{existentially conditioned equations
  (ece's)} in this signature. Here, an existentially conditioned
equation is a sentence of the form $\Land_{i=1}^n\IsDef\alpha_i\impl
\beta\exeq\gamma$, where $\beta$, $\gamma$, and the $\alpha_i$ are
terms formed from the basic operations, application, $\lambda$\dash
abstraction, and typed variables from a given context, and
$\IsDef\alpha$ abbreviates the formula $\alpha\exeq\alpha$, which
states that the term $\alpha$ is defined. Note that since the
higher-order internal logic recalled in Section~\ref{sec:intlog} is
defined through equality, a theory may alternatively be seen as having
axioms using the full power of the internal logic.

In the correspondence between categories and theories, one associates
to every pccc with equality, $\BC$, an \emph{internal language}
$\Th(\BC)$ which has the objects of $\BC$ as basic types and the
partial morphisms as operations, as well as all ece's expressed in
this language which hold in $\BC$ as axioms. Conversely, one
associates to every theory $\CT$ in the partial $\lambda$\dash
calculus with equality a pccc with equality, $\Cl(\CT)$, the
\emph{classifying category} of $\CT$. The objects of $\Cl(\CT)$ are
pairs $(\Gamma\obsep\phi)$ consisting of a finite \emph{context}
$\Gamma=(x_1:s_1;\dots;x_n:s_n)$ of variables $x_i$ with assigned
types $s_i$ and a formula (i.e.\ by the correspondence between
predicates and partial functions discussed above just a definedness
assertion) $\phi$ in context $\Gamma$. As (unlike
in~\cite{Schroder06a}) we include explicit product types in the
language, we often assume that objects are just of the form
$(x:s\obsep\phi)$, in order to avoid cluttering the notation. Such an
object is thought of as the \emph{subtype} of $s$ determined by the
property $\phi$. Morphisms
$\sigma:(\Gamma\obsep\phi)\to(\Delta\obsep\psi)$ are (type-correct)
substitutions of the variables in $\Delta$ by terms in context
$\Gamma$ such that $\phi$ entails $\psi\sigma$ as well as definedness
of $\sigma(x)$ for every variable $x$ in $\Delta$; morphisms are taken
modulo provable equality of terms under $\phi$. In the simplified case
where $\Delta$ is of the form $(y:t)$, morphisms can just be regarded
as being represented by single terms.

The central facts establishing that the above correspondence is
actually an equivalence are that 
\begin{enumerate}[$\bullet$]
\item the pccc $\BC$ is equivalent to $\Cl(\Th(\BC))$, and
\item the theory $\Th(\Cl(\CT))$ is a conservative extension of $\CT$.
\end{enumerate}
When reasoning about a pccc with equality, $\BC$, one may thus assume
that $\BC$ is actually of the form $\Cl(\Th(\BC))$, i.e.\ freely move
back and forth between logical and categorical arguments, and in
particular construct objects in $\BC$ as subtypes of types formed from
$\BC$\dash objects. It is therefore helpful to recall how some
important categorical concepts are reflected in the internal logic:
\begin{enumerate}[$\bullet$]
\item Composition is chaining of substitutions.
\item The product of types $(\Gamma\obsep\phi)$ and
  $(\Delta\obsep\psi)$, where the contexts $\Gamma$ and $\Delta$ are
  w.l.o.g.\ disjoint, is $(\Gamma;\Delta\obsep\phi\land\psi)$.
 \item Identities and product projections are just variables.
 \item The equaliser of two morphisms
   $(x:s\obsep\phi)\to(y:t\obsep\psi)$ given by terms $\alpha$,
   $\beta$ in context $x:s$ is the type
   $(x:s\obsep\phi\land\alpha\exeq\beta)$. 
\end{enumerate}
The constructions of initial datatypes and final
process types in Sect.~\ref{sec:cons} will be based on this
principle. They will, by the above equivalence, amount simultaneously
to conservativity results in \HasCASL and to existence theorems for
datatypes and process types in the categorical semantics. In the
latter incarnation, they apply in particular to
quasitoposes~\cite{Wyler91}, which may be defined as finitely
cocomplete pccc's with equality. This class of categories is
technically related to toposes, the essential difference being that
the internal logic of a topos speaks about \emph{all} subobjects,
while the internal logic of a quasitopos speaks only about the
regular subobjects (or more formally that the classifier of a topos
classifies all subobjects, and that of a quasitopos only the regular
subobjects).
\begin{rem}
  Quasitoposes have a further, first order internal logic which is
  based on the full subobject fibration. Throughout this work, we use
  the term `internal logic of a quasitopos', or more generally of a
  pccc with equality, to refer to the higher order internal logic
  based on the regular subobject fibration.
\end{rem}

\noindent As mentioned above, the range of examples is much broader in
the case of quasitoposes; e.g.\ there are many interesting non-trivial
concrete quasitoposes over $\Set$, while concrete toposes over $\Set$
are always full subcategories of $\Set$. Intuitively, quasitoposes
support a distinction between `maps', i.e.\ functional relations, and
`morphisms', i.e.\ functions, while the two concepts coincide in
toposes. Similarly, quasitoposes distinguish between partial maps,
designated below by the symbol $\pfun$, and single-valued
relations. In the internal logic, the difference is captured precisely
by the fact that toposes admit unique choice, while quasitoposes do
not. Objects $A$ in a quasitopos (or in a pccc with equality) that do
admit unique choice in the sense described in Sect.~\ref{sec:intlog}
are called \emph{coarse}. Explicitly, $A$ is coarse iff there exists a
function $c$ from the type $\Sg(A)$ of singleton subsets of $A$ to $A$
such that $c(p)$ is in $p$ (and hence $p=\{c(p)\}$) for every
$p:\Sg(A)$; in this case, the unique description term $\iota
x:A\sbullet\phi$ can be defined as $c\ (\lambdaTerm{x:A}{\phi})$.

To give the reader a basic feeling for the above issues, we recall one
of the simplest examples of a non-trivial set-based quasitopos, the
category $\ReRe$ of reflexive relations~\cite{AdamekHerrlich90}. The
objects of $\ReRe$ are pairs $(X,R)$ with $R$ a reflexive relation on
the set $X$, and morphisms $f:(X,R)\to(Y,S)$ are relation-preserving
maps $f:X\to Y$, i.e.\ $f(x)Sf(y)$ whenever $xRy$. We say that $(X,R)$
is \emph{discrete} if $R$ is equality, and \emph{indiscrete} if
$R=X\times X$. The coarse objects of $\ReRe$ are precisely the
indiscrete objects. The category $\ReRe$ has a natural numbers object,
i.e.\ an initial algebra for the functor $\argument+1$, namely the
discrete structure on the set of natural numbers. In particular, the
natural numbers object fails to be coarse, i.e.\ does not support
unique choice.

\section{Datatypes in \secHasCASL}\label{sec:datatypes}

\noindent \HasCASL supports recursive datatypes in the same style as
in \CASL~\cite{CASL-UM,CASL-RM}. To begin, an unconstrained datatype
$t$ is declared along with its constructors $c_i:t_{i1}\tfun\dots\tfun
t_{ik_i}\tfun t$ (where the function arrows $\tfun$ and $\pfun$ are
right associative) by means of the keyword \Type in the form
\begin{hetcasl}
\> \KW{type} \=  \Ax{::=} \= \Id{$c_1$} \Id{$t_{11}$} \dots \Id{$t_{1k_1}$} 
\AltBar{} 
\dots \AltBar{} 
\Id{$c_n$} \Id{$t_{n1}$} \dots \Id{$t_{nk_n}$} 
\end{hetcasl}
\noindent (mutually recursive types are admitted as well, but omitted
from the presentation for the sake of readability; their handling
requires essentially no more than adding more indices).  Here, $t$ is
a pattern of the form $C\ a_1\ \dots\ a_r$, $r\geq 0$, where $C$ is
the type constructor (or type if $r=0$) being declared and the $a_i$
are type variables. The $t_{ij}$ are types whose formation may involve
$C$, the type variables~$a_i$, and any types declared in the
\emph{local environment}, i.e.\ the context of preceding declarations.
Optionally, selectors $\mi{sel}_{ij}:t\pfun t_{ij}$ may be declared by
writing $(\mi{sel}_{ij}:?t_{ij})$ in place of $t_{ij}$. All this is
syntactic sugar for the corresponding declarations of types and
operations, and equations stating that selectors are left inverse to
constructors.

Datatypes may be qualified by a preceding \Free\ or \Generated. The
\Generated\ constraint introduces an induction axiom; this corresponds
roughly to term generatedness (`no junk'). The \Free\ constraint (`no
junk, no confusion') instead introduces an implicit fold operator,
which implies both induction and a primitive recursion principle. If
one of these constraints is used, then recursive occurrences (in
the~$t_{ij}$) of $C$ are restricted to the pattern $t=C\ a_1\ \dots\
a_r$ appearing on the left hand side; i.e.\ \HasCASL does not support
polymorphic recursion. If a \Free constraint is used, then
additionally recursive occurrences of $t$ are required to be strictly
positive w.r.t.\ function arrows, i.e.\ occurrences in the argument
type of a function type are forbidden. We omit a detailed discussion
of generatedness
constraints~\cite{SchroderMossakowski08HCOverview}. The semantics of
freeness constraints is defined in more detail as follows.

Standardly, initial datatypes are characterised by the abovementioned
induction axioms (\emph{no junk}) and additionally by the \emph{no
  confusion} condition, stating essentially that all terms formed from
the constructors and given elements of the types in the local
environment denote distinct values. By the discussion in
Sect.~\ref{sec:semantics}, it is clear that these conditions are
insufficient in the setting of \HasCASL's internal logic: in the maps
vs.\ morphisms metaphor, they constrain only the underlying set of a
datatype, not its structure. E.g.\ in the quasitopos $\ReRe$ of
reflexive relations, the no-junk-no-confusion axioms for the datatype
of natural numbers, i.e.\ the Peano axioms, will be satisfied by any
object whose underlying set is the set of natural numbers. In
particular, one will not be able to prove a recursion principle from
the Peano axioms (which is possible under unique
choice~\cite{Paulson95}), as models of the Peano axioms in general
fail to be initial algebras.

As mentioned above, the semantics of free datatypes in \HasCASL is
therefore determined by a fold operator, i.e.\ free datatypes are
explicitly axiomatised as initial algebras.  As indicated above,
recursive occurrences of free types must be strictly positive, i.e.\
types like $L::=\Id{abs}\ (L\tfun L)$ and $L\ a::=\Id{abs}\ ((L\tfun
a)\tfun a)$ are illegal, while
\begin{hetcasl}
\> \KW{free} \KW{type} \=\Id{Tree} \Id{a} \Id{b} \Ax{::=} \=\Id{leaf} \Id{b} \AltBar{} \Id{branch} (\=\Id{a} \Ax{\to} \=\Id{Tree} \Id{a} \Id{b})
\end{hetcasl}
\noindent is allowed. Free datatypes may thus be seen as initial
algebras for functors. In the standard case, the functors in question
are polynomial functors, with multiple arguments of constructors
represented as products and alternatives represented as sums. E.g.\
the signature of the tree type above induces the functor $F_{ab}$
given by
\begin{equation*}
F_{ab} c = b + (a \to c).
\end{equation*}
%where the dependence on $b$ (but not on $a$) is also functorial. 
\noindent The general mechanism for extracting functors from datatype
declaration is explained in more detail in
Sect.~\ref{sec:bootstrap}. This mechanism relies on type classes to
ensure that user-defined type constructors appearing in constructor
arguments are actually functors. The latter will in particular be the
case if type constructors are defined as free datatypes with functorial
parameters; e.g.\ the above declaration induces a functor taking $b$ to
$\mi{Tree}\ a\ b$.

For now, we take for granted that a free datatype $t$ as in the
beginning of this section can be regarded as an initial algebra
$\alpha:F\ t\to t$ for a functor $F$. Initiality is expressed by means
of a polymorphic fold operation
\begin{equation*}
\mi{fold}:  (F\ b\to b) \to  t\to b
\end{equation*}
for $b:\TypeKind$, and an axiom stating that, for $d: F\ b\to b$,
$\mi{fold}\ d$ is the unique $F$\dash algebra morphism from $\alpha$ to
$d$, i.e.\ the unique map $f:t \tfun b$ satisfying
\begin{equation*}
d\circ(F\ f) = f\circ\alpha.
\end{equation*}
Initiality implies induction and term distinctness, i.e.\ the usual
no-junk/no-confusion conditions: term distinctness follows from the
fact that structure maps of initial algebras are isomorphisms
(Lambek's lemma); induction for a predicate $P$ on $t$ is proved by
applying $\mi{fold}$ at the type $b=(x:t\obsep P\ x)$. (The semantics
of polymorphism in \HasCASL prescribes that polymorphic operators such
as fold do have instances at
subtypes~\cite{SchroderMossakowski08HCOverview}. For polynomial
functors, the use of such instances can be circumvented; see
Remark~\ref{rem:nno} for more comments on this point.) Moreover, one
obtains a \emph{primitive recursion} principle by means of a
simultaneous recursive definition of the identity (as suggested
in~\cite{Girard1989}): The fold operation allows defining recursive
functions $f:t\to b$, where $\alpha:F\ t\to t$ is the initial datatype
for the functor $F$, using the iteration scheme, i.e.\
\begin{equation*}
  f\ (\alpha\ x)=d\ (F\ f\ x)\quad\textrm{for $x: F\ t$}
\end{equation*}
(which is just a restatement of the previous equation). One may thus
in particular define a function $g:t\to t\times b$ by
\begin{align*}
  g\ (\alpha \ x)& =(\lambdaTerm{y:F(t\times b)}
  {(\alpha\ (F\ \pi_1\ y),d\ y)})\ (F\ g\ x)\\
  & = (\alpha\ (F\ \pi_1\ (F\ g\ x)),d\ (F\ g\ x))\\
  & = (\alpha\ (F\ (\pi_1\circ g)\ x)),d\ (F\ g\ x))
\end{align*}
where $\pi_1$ denotes the first projection $\lambdaTerm{(x,z):t\times
  b}{x}$. Here, the actual body of the recursive definition is the map
$d:F(t\times b)\tfun b$. The defining equation of $g$ implies that
$(\pi_1\circ g)\ (\alpha\ x) = \alpha\ (F\ (\pi_1\circ g)\ x))$ and
hence, by uniqueness of folds, that the first component $\pi_1\circ g$
of $g$ is the identity on $t$. Therefore,
$g=\lambdaTerm{y:t}{(y,(\pi_2\circ g)\ y)}$, so that the second component
$f=\pi_2\circ g$ of $g$, where $\pi_2:t\times b\to b$ denotes the second
projection, satisfies
\begin{equation*}
  f\ (\alpha\ x)=d\ (F(\lambdaTerm{y:t}{(y,f\ y)})\ x).
\end{equation*}
Conversely, every solution $f$ of this equation yields a solution
$g=\lambdaTerm{y:t}{(y,f\ y)}$ of the iteration equation for $g$. Thus
we may define $f:t\to b$ by \emph{primitive recursive equations},
whose right hand side may depend on applications $f\ x$ of $f$ to
constructor arguments $x$ appearing in the pattern $\alpha\ x
$ on the
left hand side, as in the case of iteration, and additionally on the
constructor arguments $x$ themselves.

Since by Lambek's lemma, the structure map of an initial algebra is an
isomorphism, free datatypes $\alpha:F\ t\tfun t$ inherit a case
operator from the decomposition of $F\ t$ as a sum; such an operator
\begin{equation*}
\mi{case}\ x\ \mi{of}\ c_1\ y_{11}\ \dots\ y_{1k_1} \to f_1\   y_{11}\ \dots\ y_{1k_1} \mid
\ldots \mid c_1\ y_{n1}\ \dots\ y_{nk_n} \to f_1\   y_{n1}\ \dots\ y_{nk_n}
\end{equation*}
is provided explicitly in \HasCASL.

\begin{rem}\label{rem:non-standard}
Unlike in \CASL, the meaning of \Free\ \Type\ does not coincide with
that of the corresponding structured free extension \Free $\Group$ \Type
\dots $\EndGroup$, which would require all newly arising function types
to be also freely term generated.
\end{rem}

% \begin{rem}
% The reason for using an explicit fold operation in place of a
% combination of induction (`no junk') and term distinctness (`no
% confusion') is the absence of a unique choice operator, without which
% the existence of folds fails to be derivable from the
% no-junk-no-confusion principles. (E.g.\ in the quasitopos of
% pseudotopological spaces, the no-junk-no-confusion axioms determine the
% underlying set of an initial algebra but not its topological structure.)

\begin{exa}\label{expl:freetypes}
Consider the following free datatype definitions.
\begin{hetcasl}
\> \KW{free} \KW{type} \=\Id{List} \Id{a} \Ax{::=} \=\Id{nil} \AltBar{} \Id{cons} (\=\Id{a}; \=\Id{List} \Id{a})\\
\> \KW{free} \KW{type} \=\Id{Tree} \Id{a} \Id{b} \Ax{::=} \=\Id{leaf} \Id{a} \AltBar{} \Id{branch} (\=\Id{b} \Ax{\to} \=\Id{List} (\=\Id{Tree} \Id{a} \Id{b}))
\end{hetcasl}
\noindent The declaration of $\mi{List}\ a$ induces the standard fold
operation for lists. Moreover, the type class mechanism (see
Section~\ref{sec:bootstrap}) recognises automatically that the type
constructor $\mi{List}$ is a functor, and in particular generates the
standard $\mi{map}$ operation. For $\mi{Tree}$, we obtain a polymorphic
fold operation
\begin{equation*}
\mi{fold}:  (a\to c) \to 
((b \to \mi{List}\ c)\to c) \to \mi{Tree}\ a\ b \to c.
\end{equation*}
This operation is axiomatised as
being uniquely determined by the equations
\begin{equation*}
\mi{fold}\ f\ g\ (\mi{leaf}\ x)  = f\ x\quad\textrm{and}\quad
\mi{fold}\ f\ g\ (\mi{branch}\ s)  = 
  g\ (\mi{map}\  (\mi{fold}\ f\ g)\circ s).
\end{equation*}
\end{exa}

\section{Initiality via the Type Class Mechanism}
\label{sec:bootstrap}

\begin{myfigure}
\fbox{\parbox{\myboxwidth}{
\vspace{-10pt}
\begin{hetcasl}
\SPEC \=\SIdIndex{Functor} \Ax{=}\\
\> \VARS \=\Id{a}, \Id{b}, \Id{c} \Ax{:} \Id{Type}; \Id{x} \Ax{:} \Id{a}; \Id{f} \Ax{:} \=\Id{a} \Ax{\to} \Id{b}; \Id{g} \Ax{:} \=\Id{b} \Ax{\to} \Id{c}\\
\> \OPS \=\Ax{\_\_}\Id{comp}\Ax{\_\_} \Ax{:} \=(\=\Id{b} \Ax{\to} \Id{c}) \Ax{\times} (\=\Id{a} \Ax{\to} \Id{b}) \Ax{\to} \=\Id{a} \Ax{\to} \Id{c};\\
\>\> \Id{id} \Ax{:} \=\Id{a} \Ax{\to} \Id{a}\\
\> \Ax{\bullet} \=\Id{id} \Id{x} \Ax{=} \Id{x}\\
\> \Ax{\bullet} \=(\=\Id{g} \Id{comp} \Id{f}) \Id{x} \Ax{=} \=\Id{g} (\=\Id{f} \Id{x})\\
\> \CLASS \Id{Functor} \Ax{<} \=\Id{Type} \Ax{\to} \Id{Type}\\
\> \{\VARS \=\Id{a}, \Id{b}, \Id{c} \Ax{:} \Id{Type}; \Id{F} \Ax{:} \Id{Functor}; \Id{f} \Ax{:} \=\Id{a} \Ax{\to} \Id{b}; \Id{g} \Ax{:} \=\Id{b} \Ax{\to} \Id{c}\\
\> \OP \=\Id{map} \Ax{:} \=(\=\Id{a} \Ax{\to} \Id{b}) \Ax{\to} \=\Id{F} \Id{a} \Ax{\to} \=\Id{F} \Id{b}\\
\> \Ax{\bullet} \=\Id{map} \Id{id} \Ax{=} \=\Id{id} \Ax{:} \=\Id{F} \Id{a} \Ax{\to} \=\Id{F} \Id{a}\\
\> \Ax{\bullet} \=\Id{map} (\=\Id{g} \Id{comp} \Id{f}) \Ax{:} \=\Id{F} \Id{a} \Ax{\to} \=\Id{F} \Id{c} \Ax{=}  (\=\Id{map} \Id{g}) \Id{comp} (\=\Id{map} \Id{f})\\
\> \}\\
\> \CLASS \Id{Bifunctor} \Ax{<} \=\Id{Type} \Ax{\to} \Id{Functor}\\
\> \{\VARS \=\Id{a}, \Id{b}, \Id{c}, \Id{d} \Ax{:} \Id{Type}; \Id{F} \Ax{:} \Id{Bifunctor}; \Id{f} \Ax{:} \=\Id{a} \Ax{\to} \Id{b}; \Id{g} \Ax{:} \=\Id{b} \Ax{\to} \Id{c}; \Id{h} \Ax{:} \=\Id{c} \Ax{\to} \Id{d}\\
\> \OP \=\Id{parmap} \Ax{:} \=(\=\Id{a} \Ax{\to} \Id{b}) \Ax{\to} \=\Id{F} \Id{a} \Id{d} \Ax{\to} \=\Id{F} \Id{b} \Id{d}\\
\> \Ax{\bullet} \=\Id{parmap} \Id{id} \Ax{=} \=\Id{id} \Ax{:} \=\Id{F} \Id{a} \Id{d} \Ax{\to} \=\Id{F} \Id{a} \Id{d}\\
\> \Ax{\bullet} \=\Id{parmap} (\=\Id{g} \Id{comp} \Id{f}) \Ax{:}
\=\Id{F} \Id{a} \Id{d} \Ax{\to} \=\Id{F} \Id{c} \Id{d} \Ax{=}
(\=\Id{parmap} \Id{g}) \Id{comp} (\=\Id{parmap} \Id{f});\\
\> \Ax{\bullet} (\=\Id{parmap} \Id{f}) \Id{comp} (\=\Id{map}
\Id{h}) \Ax{:} \=\Id{F} \Id{a} \Id{c} \Ax{\to} \=\Id{F} \Id{b}
\Id{d} \Ax{=} 
(\=\Id{map} \Id{h}) \Id{comp} (\=\Id{parmap} \Id{f})\\
\> \}

\end{hetcasl}\hfill
\vspace{-20pt}
}}
\caption{\HasCASL specification of functors}
\label{fig:functors}
\end{myfigure}

\noindent The concept of free datatype described in the previous
section may be regarded as bootstrapped, i.e.\ as being a \HasCASL
library equipped with built-in syntactic sugar rather than a basic
language feature. The crucial point here is that \HasCASL's type class
mechanism allows talking about functorial signatures, algebras for a
functor, and algebra homomorphisms. 
% The specifications shown below are
% real \HasCASL specifications, parsed and prettyprinted using the
% Heterogeneous Tool Set (Hets)~\cite{MossakowskiEA07}.

Figure~\ref{fig:functors} shows the constructor class of
functors. Mutually recursive or parametrised datatypes require $n$\dash ary
functors for $n\in\Nat$, and in fact occasionally type constructors
which are functorial only in some of their arguments; since \HasCASL
does not feature dependent classes, the corresponding classes need to be
specified one by one, as exemplified in Fig.~\ref{fig:functors} by a
specification of bifunctors. This is not a problem in practice, as
typically only small values of $n$ are needed; the specification of
bifunctors illustrates how $n+1$\dash ary functors can be specified
recursively in terms of $n$\dash ary functors.

\begin{rem}
One might envision a single specification of functors of arbitrary
finite arity by abuse of syntax, as follows: declare a class
$\mi{Typelist}$ and type constructors $\mi{Nil}:\mi{Typelist}$,
$\mi{Cons}:\TypeKind\to\mi{Typelist}\to\mi{Typelist}$, and define
$\mi{Functor}$ as a subclass of $\mi{Typelist}\to\TypeKind$. (Undesired
semantic side effects may be eliminated by specifying the types
$\mi{Nil}$, $\mi{Cons}\ a\ \mi{Nil}$ etc.\ to be singletons.)  Similar
tricks work in Haskell~\cite{KiselyovEA04} but rely  on
multi-parameter type classes, which are currently excluded from the
\HasCASL design.
\end{rem}

\begin{myfigure}
\fbox{\parbox{\myboxwidth}{
\vspace{-10pt}
\begin{hetcasl}
\SPEC \=\SIdIndex{PolyFunctors} \Ax{=}\\
\> \SId{Functor}\\
\THEN \=\CLASSES \=\Id{PolyFunctor} \Ax{<} \Id{Functor};\\
\> \> \=\Id{PolyBifunctor} \Ax{<} \=\Id{Type} \Ax{\rightarrow} \Id{PolyFunctor};\\
\>\> \Id{PolyBifunctor} \Ax{<} \Id{Bifunctor}\\
\> \VARS \=\Id{F}, \Id{G} \Ax{:} \Id{PolyFunctor}; \Id{H}, \Id{K} \Ax{:} \Id{PolyBifunctor}; \=\Id{a}, \Id{b}, \Id{c} \Ax{:} \Id{Type}\\
\> \TYPE \=\Id{Sum} \=\Id{b} \Id{c} \Ax{::=} \=\Id{inl} \Id{b} \AltBar{} \Id{inr} \Id{c}\\
\> \VARS \=\Id{f} \Ax{:} \=\Id{b} \Ax{\rightarrow} \Id{a}; \Id{g} \Ax{:} \=\Id{c} \Ax{\rightarrow} \Id{a}; \Id{h} \Ax{:} \=\Id{Sum} \Id{b} \Id{c} \Ax{\rightarrow} \Id{a}\\
\> \OP \=\Id{sumcase} \Ax{:} \=(\=\Id{b} \Ax{\rightarrow} \Id{a}) \Ax{\rightarrow} \=(\=\Id{c} \Ax{\rightarrow} \Id{a}) \Ax{\rightarrow} \=\Id{Sum} \Id{b} \Id{c} \Ax{\rightarrow} \Id{a}\\
\> \Ax{\bullet} \=\Id{h} \Ax{=} \=\Id{sumcase} \Id{f} \Id{g} \\
\>\> \Ax{\Leftrightarrow} \=\Ax{\forall} \Id{x} \Ax{:} \Id{b}; \Id{y} \Ax{:} \Id{c} \Ax{\bullet} \=\Id{h} (\=\Id{inl} \Id{x}) \Ax{=} \=\Id{f} \Id{x} \Ax{\wedge} \=\Id{h} (\=\Id{inr} \Id{y}) \Ax{=} \=\Id{g} \Id{y};\\
\> \TYPES \=\Id{Fst} \Id{a} \Id{b} \Id{\Ax{:}\Ax{=}} \Id{a};\\
\>\> \Id{Snd} \Id{a} \Id{b} \Id{\Ax{:}\Ax{=}} \Id{b};\\
\>\> \Id{Id} \Id{b} \Id{\Ax{:}\Ax{=}} \Id{b};\\
\>\> \Id{ProdF} \Id{F} \Id{G} \Id{b} \Id{\Ax{:}\Ax{=}} \=\Id{F} \Id{b} \Ax{\times} \=\Id{G} \Id{b};\\
\>\> \Id{ProdBF} \Id{H} \Id{K} \Id{b} \Id{c} \Id{\Ax{:}\Ax{=}} \=\Id{H} \Id{b} \Id{c} \Ax{\times} \=\Id{K} \Id{b} \Id{c};\\
\>\> \Id{SumF} \Id{F} \Id{G} \Id{b} \Id{\Ax{:}\Ax{=}} \=\Id{Sum} (\=\Id{F} \Id{b}) (\=\Id{G} \Id{b});\\
\>\> \Id{SumBF} \Id{H} \Id{K} \Id{b} \Id{c} \Id{\Ax{:}\Ax{=}} \=\Id{Sum} (\=\Id{H} \Id{b} \Id{c}) (\=\Id{K} \Id{b} \Id{c})\\
\> \TYPES \=\Id{Fst} \Id{a}, \Id{Id}, \Id{ProdF} \=\Id{F} \Id{G}, \Id{SumF} \=\Id{F} \Id{G}  \Ax{:} \Id{PolyFunctor};\\
\>\> \Id{Fst}, \Id{Snd},  \Id{ProdBF} \=\Id{H} \Id{K}, \Id{SumBF} \=\Id{H} \Id{K}  \Ax{:} \Id{PolyBifunctor}\\
\> \VAR \Id{k} \Ax{:} \=\Id{a} \Ax{\rightarrow} \Id{b}\\
\> \Ax{\bullet} \=(\=\Id{map} \Id{k} \Ax{:} \=\Id{SumF} \Id{F} \Id{G} \Id{a} \Ax{\rightarrow} \=\Id{SumF} \Id{F} \Id{G} \Id{b}) \\
\>\>\>\> \Ax{=} \=\Id{sumcase} (\=\Id{inl} \Id{comp} \=\Id{map} \Id{k}) (\=\Id{inr} \Id{comp} \=\Id{map} \Id{k})\\
\> \Ax{\bullet} \=(\=\Id{parmap} \Id{k} \Ax{:} \=\Id{SumBF} \Id{H} \Id{K} \Id{a} \Id{c} \Ax{\rightarrow} \=\Id{SumBF} \Id{H} \Id{K} \Id{b} \Id{c}) \\
\>\>\>\> \Ax{=} \=\Id{sumcase} (\=\Id{inl} \Id{comp} \=\Id{parmap} \Id{k}) (\=\Id{inr} \Id{comp} \=\Id{parmap} \Id{k})\\
\> \Ax{\bullet} \dots {\small{}\KW{\%\%} definitions of map and parmap for the other cases}
\end{hetcasl}\hfill
\vspace{-20pt}
}}
\caption{\HasCASL specification of polynomial functors}
\label{fig:polyfunctors}
\end{myfigure}

\noindent For purposes of conservativity of datatype declarations, the
class of polynomial functors (bifunctors etc.), shown in
Fig.~\ref{fig:polyfunctors}, plays an important role. An $n$\dash ary
functor is polynomial if it can be generated from projection functors
(the identity functor if $n=1$) and constant functors by taking finite
sums and products. These operations, and similar constructions in
Fig.~\ref{fig:algebras}, are defined as \emph{type synonyms},
i.e.\ as mere abbreviations of existing types.\footnote{Consequently,
  the specifications, while correct according to the \HasCASL language
  definition, fail to pass the static analysis in the present version
  of the heterogeneous tool set Hets~\cite{MossakowskiEA07}, as type
  synonyms are currently immediately expanded and $\beta$\dash reduced;
  this will be remedied in future versions of the tool.
  In~\cite{Schroder07a}, we have used type declarations with explicit
  constructors as a workaround in place of type synonyms; for purposes
  of the present work, we have given preference to readability of
  specifications.} The obvious definitions of the $\mi{map}$ and
$\mi{parmap}$ operations are omitted for most of the functors
introduced in Fig.~\ref{fig:polyfunctors}, except in the case of
sums. Note that \HasCASL does not provide a way to exclude unwanted
(`junk') further instance declarations for the class
$\mi{PolyFunctor}$, i.e.\ to say that the class is generated by the
given generic instances. As in Fig.~\ref{fig:functors}, we show only
the specification for functors of arity at most $2$; the extension to
higher arities is obvious.

In Fig.~\ref{fig:algebras}, we present a specification of algebras for a
functor. The set of algebra structures for a functor $F$ on a type $a$
is given by the type constructor $\mi{Alg}$, which depends on both $F$
and $a$ and thus has the profile $\mi{Functor}\to\TypeKind\to\TypeKind$;
it is given as a type synonym for the type $F\ a\to a$. Similarly, the
type constructor $\mi{AlgMor}$ for algebra morphisms depends on $F$ and
types $a$, $b$ forming the carriers of the domain and the codomain,
respectively. Algebra morphisms are treated as triples consisting of two
algebra structures and a map between the carriers, thus circumventing
the absence of dependent types --- such as the `type' of algebra
morphisms between algebras \(alpha\) and \(beta\) --- in \HasCASL (these
can be conservatively added to the language~\cite{Schroder04}, however
at the price of making type checking undecidable).

\begin{myfigure}
\fbox{\parbox{\myboxwidth}{
\vspace{-10pt}
\begin{hetcasl}
\SPEC \=\SIdIndex{Algebra} \Ax{=} \SId{PolyFunctors}\\
\THEN \=\VARS \=\Id{F} \Ax{:} \Id{Functor}; \Id{a}, \Id{b} \Ax{:} \Id{Type}\\
\> \TYPE \=\Id{Alg} \Id{F} \Id{a} \Id{\Ax{:}\Ax{=}}\=\Id{F} \Id{a} \Ax{\rightarrow} \Id{a}\\
\> \OP \=\Ax{\_\_}\Ax{::}\Ax{\_\_}\Ax{\to}\Ax{\_\_} \Ax{:} \=\Id{Pred} (\=(\=\Id{a} \Ax{\rightarrow} \Id{b}) \Ax{\times} (\=\Id{Alg} \Id{F} \Id{a}) \Ax{\times} (\=\Id{Alg} \Id{F} \Id{b}))\\
\> \VARS \=\Id{f} \Ax{:} \=\Id{a} \Ax{\rightarrow} \Id{b}; \Id{alpha} \Ax{:} \=\Id{Alg} \Id{F} \Id{a}; \Id{beta} \Ax{:} \=\Id{Alg} \Id{F} \Id{b}\\
\> \Ax{\bullet} \=(\=\Id{f} \Ax{::} \Id{alpha} \Ax{\to} \Id{beta})
\Ax{\Leftrightarrow}  (\=\Id{beta} \Id{comp} (\=\Id{map} \Id{f})) \Ax{=}
(\=\Id{f} \Id{comp} \Id{alpha})\\
\> \TYPE \=\Id{AlgMor} \Id{F} \Id{a} \Id{b} \Ax{=} 
\{(\=\Id{f}, \Id{alpha}, \Id{beta}) \Ax{:} (\=\Id{a}
\Ax{\rightarrow} \Id{b}) \Ax{\times} \Id{Alg} \Id{F} \Id{a} \Ax{\times}
\=\Id{Alg} \Id{F} \Id{b} \Ax{\bullet} \\
\>\>\>\>\Id{f} \Ax{::} \Id{alpha} \Ax{\to} \Id{beta} \}\\
\> \CLASSES \Id{DTFunctor} \Ax{<} \Id{Functor};  
\Id{PolyFunctor} \Ax{<} \Id{DTFunctor}\\
\> \{\VARS \=\Id{F} \Ax{:} \Id{DTFunctor}; \Id{a} \Ax{:} \Id{Type}\\
\> \TYPE \Id{InitialCarrier} \Id{F}\\
\> \OPS \=\Id{initialAlg} \Ax{:} \=\Id{Alg} \Id{F} (\=\Id{InitialCarrier} \Id{F});\\
\>\> \Id{ifold} \Ax{:} \=\Id{Alg} \Id{F} \Id{a} \Ax{\rightarrow}
\=\Id{InitialCarrier} \Id{F} \Ax{\rightarrow} \Id{a}\\
\> \VARS \=\Id{alpha} \Ax{:} \=\Id{Alg} \Id{F} \Id{a}; \Id{g} \Ax{:} \=\Id{InitialCarrier} \Id{F} \Ax{\rightarrow} \Id{a};\\
\> \Ax{\bullet} \=(\=\Id{g} \Ax{::} \Id{initialAlg} \Ax{\to} \Id{alpha}) \Ax{\Leftrightarrow} \=\Id{g} \Ax{=} \=\Id{ifold} \Id{alpha};\\
\> \}\\
\> \VAR \Id{G} \Ax{:} \Id{PolyBifunctor}\\
\> \TYPE \=\Id{ParamDT} \Id{G} \Id{a} \Id{\Ax{:}\Ax{=}} \=\Id{InitialCarrier} (\=\Id{G} \Id{a})\\
\> \TYPE \=\Id{ParamDT} \Id{G} \Ax{:} \Id{DTFunctor}\\
\> \VARS \=\Id{l} \Ax{:} \=\Id{ParamDT} \Id{G} \Id{a}; \Id{b} \Ax{:} \Id{Type}; \Id{f} \Ax{:} \=\Id{a} \Ax{\rightarrow} \Id{b}\\
\> \Ax{\bullet} \=\Id{map} \Id{f} \Id{l} \Ax{=} \=\Id{ifold} (\=\Id{initialAlg} \Id{comp} \=\Id{parmap} \Id{f}) \Id{l}

\end{hetcasl}\hfill
\vspace{-20pt}
}}
\caption{\HasCASL specification of initial algebras}
\label{fig:algebras}
\end{myfigure}

Initial algebras are then specified by means of two operations: a type
constructor $\mi{InitialCarrier}$ that assigns to a functor the
carrier set of its initial algebra, and a polymorphic constant
$\mi{initialAlg}$ which represents the structure map of an initial
algebra for $F$ on this carrier. Initiality of this algebra is
specified by means of an explicit fold operation, called $\mi{ifold}$
in the specification. As initial algebras will exist only for some
functors, the abovementioned operations are defined only on a subclass
$\mi{DTFunctor}$ (`datatype functor') of $\mi{Functor}$. We declare
the class $\mi{PolyFunctor}$ (Fig.~\ref{fig:polyfunctors}) to be a
subclass of $\mi{DTFunctor}$, thus stating that all polynomial
functors have initial algebras as proved in Sect.~\ref{sec:cons}; due
to possible junk in the class $\mi{PolyFunctor}$ (see above), this is
consistent but non-conservative. Moreover, we state that initial
algebras depend functorially on parameters in the case of polynomial
bifunctors and that the arising functor again has an initial algebra
(as nested recursion may be coded by mutual recursion in the standard
way~\cite{Gunter93}) by defining a type constructor $\mi{ParamDT}$
which maps a polynomial bifunctor $G$ to the functor that takes a type
$a$ to the initial algebra of the polynomial functor $G\ a$, and by
declaring $\mi{ParamDT\ G}$ to be an instance of $\mi{DTFunctor}$.

\begin{rem}
  Note that functors induced by parametrised initial datatype
  declarations are declared as instances of $\mi{DTFunctor}$ in
  Fig.~\ref{fig:algebras} only if the signature functor is
  polynomial. It is not possible to extend this mechanism to arbitrary
  parametrised datatypes, as the corresponding functors need not
  themselves have initial algebras. As a simple example, consider the
  declaration
  \begin{hetcasl}
    \> \VAR \Id{a} \Ax{:} \Id{Type}\\
\> \KW{free} \KW{type} \=\Id{C} \Id{a} \Ax{::=} \Id{abs} (\=\Id{Pred} \Id{a})\\

  \end{hetcasl}
  \noindent which defines $C\ a$ as the initial algebra of the functor
  $F$ given by $F\ a \ b=\mi{Pred}\ a$, i.e.\ $F$ takes the powerset
  of its first argument and ignores the second. Thus, $C\ a$ is
  isomorphic to $\mi{Pred}\ a$ and hence, by Russell's paradox, the
  functor $C$ does not have an initial algebra.
\end{rem}

%\begin{rem}
%The specifications above illustrate that \HasCASL
%specifications will usually make do without specification
%parametrisation, since this can often be better expressed by means of
%the \HasCASL type class mechanism. A fundamental exception to this
%principle is formed by specifications which are parametrised by classes,
%which however may be expected to be a rare case.
%\end{rem}

\noindent We conclude with a brief description of how the data above
are generated by the static analysis of actual \HasCASL
specifications. The functor $F$ associated to the declaration of a
datatype $t$ as in the beginning of Sect.~\ref{sec:datatypes} is a sum
of $n$ functors $F_i$, one for each constructor $c_i$; the functor
$F_i$, in turn, is a product of $k_i$ functors $F_{ij}$, corresponding
to the $t_{ij}$. The $t_{ij}$ are, by the restrictions laid out in
Sect.~\ref{sec:datatypes}, inductively generated from types in the
local environment, $t=C\ a_1\ \dots\ a_n$, and the type variables
$a_i$ by taking products, exponentials $s\to t$ or $s\pfun t$, where
$s$ is a type formed from the $a_i$ and the local environment, and
applications $D\ s_1\ \dots\ s_l$ of type constructors from the local
environment, the latter subject to the restriction that if $s_i$
contains a recursive occurrence of $t$, then the dependence of $D$ of
its $i$\dash th argument must be functorial. The latter property is tracked
by means of the type class mechanism; in particular, instances of
$\mi{Functor}$ are generated automatically for parametrised datatypes
such as the type $\mi{List}\ a$ of Example~\ref{expl:freetypes}. Given
this format of the $t_{ij}$, it is straightforward to associate a
functor to each $t_{ij}$ (using further generic instances of
$\mi{Functor}$, in particular exponentials and closure under functor
composition). Finally, an instance $F:\mi{DTFunctor}$ is generated. If
this instance is already induced by the generic instances shown in
Fig.~\ref{fig:algebras}, then the datatype declaration is guaranteed
to be conservative (see Section~\ref{sec:cons}); otherwise,
conservativity and in fact consistency of the datatype declaration
becomes the responsibility of the user. This happens in particular
when constructor arguments involve either type constructors from the
local environment which are not themselves declared as initial
datatypes or exponentiation with types from the local
environment. Whether or not datatype declarations are conservative in
the latter case, which in particular includes the case of
\emph{infinite branching}, remains an open problem; under unique
choice, declarations of infinitely branching datatypes are
conservative~\cite{MoerdijkPalmgren00,Paulson97,BerghoferWenzel99}.
If $F$ is moreover of the class $\mi{PolyBifunctor}$ (or a
corresponding class of functors of higher arity), then an instance
$C:\mi{DTFunctor}$ is generated.

Using the $\mi{sumcase}$ operation of Fig.~\ref{fig:polyfunctors}, one
can gather the constructors of $t$ into a structure map $c:F\ t\to t$;
the freeness constraint then translates into the declaration of a
two-sided inverse $g$ of $\mi{ifold}\ c$. The fold operation on $t$ is
obtained as $\mi{fold}\ \alpha=(\mi{ifold}\ \alpha)\circ g$.

\section{Process Types in \HasCASL}\label{sec:cotypes}

\noindent Although process types in the style of \CoCASL, so-called
\emph{cotypes}~\cite{MossakowskiEA04}, are not presently included in
the \HasCASL design, the results of the previous section indicate that
cotypes could be integrated seamlessly into \HasCASL. A cotype is a
syntactic representation of a coalgebra for a signature
functor. Cotypes are declared in a similar style as types; the crucial
difference is that, while selectors are optional in a datatype, they
are mandatory in a cotype, as they constitute the actual structure map
of the coalgebra, and constructors are optional. Thus, the core of
a cotype declaration has the form
\begin{hetcasl}
  \> \KW{cotype} \Id{t} \Ax{::=} (\Id{$\mi{sel}_{11}$} \Ax{:} \Id{$t_{11}$}; 
  \dots; \Id{$\mi{sel}_{1k_1}$} \Ax{:} \Id{$t_{1k_1}$})
  \AltBar{}
  \dots
  \AltBar{}
  (\Id{$\mi{sel}_{n1}$} \Ax{:} \Id{$t_{n1}$}; 
  \dots; 
  \Id{$\mi{sel}_{nk_n}$} \Ax{:} \Id{$t_{nk_n}$})
\end{hetcasl}
\noindent where $t$ is a pattern $C\ a_1\ \dots\ a_n$ consisting of a
newly declared type constructor $C$ and type variables
$a_1,\dots,a_n$. A cotype induces axioms guaranteeing that the domains
of selectors in the same alternative agree, and that the domains of
all alternatives form a coproduct decomposition of the cotype. Thus,
e.g.\ the models of the cotype
\begin{hetcasl}
  \> \KW{cotype} \Id{Proc} \Ax{::=} (\Id{out} \Ax{:?} \Id{a};
  \Id{next} \Ax{:?} \Id{Proc}) 
  \AltBar{}
  (\Id{spawnl}, \Id{spawnr} \Ax{:} \Id{Proc})
\end{hetcasl}
\noindent are coalgebras for the functor $F$ given by $FX=a\times X +
X\times X$.  The semantics of cotypes, in particular \emph{cofree}
(i.e.\ final) cotypes, builds on a dual of the specification of
algebras (Fig.~\ref{fig:algebras}), where the type of algebras is
replaced by a type $\mi{Coalg}\ F\ a:=a\to F\ a$, the definition of
homomorphisms is correspondingly modified, and initiality is replaced
by finality, i.e.\ unique existence of morphisms \emph{into} the final
coalgebra given by an $\mi{unfold}$ operation. For cofree cotypes, the
codomains of the selectors are, as in the case of initial datatype,
required to depend functorially on $t$; of course, this will not in
general guarantee existence of final coalgebras. The extraction of
functors from cotype signatures is analogous to the case of types as
explained in Sect.~\ref{sec:bootstrap}, with two differences:
\begin{enumerate}[$\bullet$]
\item the class of functors that admit final coalgebras contains a
  generalised class of polynomial functors that allows replacing
  identity functors by exponentiation with constant exponents (see
  Section~\ref{sec:cons});
\item unlike for free types, functors arising from cofree cotype
  declarations, even for polynomial functors, typically do not have
  final coalgebras.
\end{enumerate}
We omit the discussion of cogeneratedness of cotypes.

The only subtle point in the matching between cotype declarations in
\HasCASL and coalgebras is that the conditions imposed in \CoCASL to
ensure that a cotype $t$ with associated functor $F$ decomposes as a
disjoint union of the domains of the selectors would in \HasCASL be
insufficient to guarantee existence of a single structure map $t\to F\
t$, the point being, again, the absence of unique choice. As indicated
above, we thus impose, instead of just disjointness and joint
exhaustiveness of the domains, that the cotype is the coproduct of the
domains, by introducing a polymorphic partial case operation similar
to the $\mi{sumcase}$ operation of Fig.~\ref{fig:polyfunctors}. E.g.\
for the cotype $\mi{Proc}$ above and $f,g:\mi{Proc}\pfun a$,
$\mi{case}\ f\ g=h:\mi{Proc}\to a$ is defined whenever the domains of
$f$ and $g$ equal the domains of $\mi{out}$ and $\mi{spawnl}$,
respectively, and in this case $h$ extends $f$ and $g$. (Under unique
choice, $\mi{case}\ f\ g$ is definable as
$\lambdaTerm{p:\mi{Proc}}{\iota x:a\sbullet x=f(p)\lor x=g(p)}$.)

\section{Conservativity of Datatypes and Process Types}\label{sec:cons}

\noindent Free datatypes in \HasCASL are not necessarily conservative
extensions of the local environment. Already the naturals may be
non-conservative: as discussed in Sect.~\ref{sec:hascasl},
conservative extensions can only introduce names for entities already
in the present signature, and a given model might interpret all types
as finite sets. This problem arises already in standard HOL, where the
construction of initial datatypes~\cite{Paulson97,BerghoferWenzel99}
is based on the naturals. The constructions given
in~\cite{Paulson97,BerghoferWenzel99} make heavy use of unique choice,
so that the question arises whether similar constructions are possible
in \HasCASL. Below, we answer this question in the affirmative for the
case of finitely branching datatypes; it remains open for the
infinitely branching case. By the equivalence of \HasCASL with the
internal logic of partial cartesian closed categories with equality
(Sect.~\ref{sec:semantics}), our results extend to pccc's with
equality and finite coproducts, and hence in particular to
quasitoposes.

To begin, we fix the required additional infrastructure. As seen
above, already the construction of signature functors for standard
datatypes requires finite sums. These are specified
(non-conservatively) in \HasCASL by declaring a type constructor
$\mi{Sum}$ as in Fig.~\ref{fig:polyfunctors}, and moreover an initial
type $\mi{Zero}$ which is specified as having a function
$\mi{zero}:\mi{Zero}\to a$ into every type $a$ and satisfying the
axiom $\forall x:\mi{Zero}.\,\bot$, where we use $\top$ and $\bot$ to
denote truth and falsity, respectively. This axiom means that the type
$\mi{Zero}$ is uninhabited, and implies that for every type $a$,
$\mi{zero}$ is the only function $\mi{Zero}\to a$. Below, we denote
sums by $+$ and the initial type by $0$ in the interest of
readability, with injection functions $\mi{inl}:A\to A+B$ and
$\mi{inr}:B\to A+B$ as usual; moreover, we use the standard
$\mi{case}$ notation as discussed in the case of initial datatypes in
Sec.~\ref{sec:datatypes}, and denote the unit type $\UnitType$ by
$1$. We write $\mi{Bool}$ for the sum type $1+1$, denoting the
associated case operator as $\mi{if}-\mi{then}-\mi{else}$ and the
terms $\mi{inl}\ ()$ and $\mi{inr}\ ()$ as $\mi{true}$ and
$\mi{false}$, respectively. We refer to the extension of the partial
$\lambda$-calculus with equality by $+$, $0$, and the associated
operators and axioms as the \emph{partial $\lambda$-calculus with
  equality and sums}.
% We define extraction functions $\mi{outl}:a+b\pfun a$,
% $\mi{outr}:a+b\pfun b$ by $\mi{outl}\ z = \mi{case}\ z\ \mi{of}\
% \mi{inl}\ x \to x\mid \mi{inr}\ y \to \mi{bot}$, analogously for
% $\mi{outr}$.

\begin{rem}
  In a cartesian closed category, finite coproducts are always
  internal in the sense that copairing is embodied in an operation
  ($\mi{sumcase}$ in the above notation) which satisfies the relevant
  laws (Fig.~\ref{fig:polyfunctors}) internally~\cite{Pitts00}. Thus,
  the internal language $\Th(\BC)$ of a pccc $\BC$ with equality and
  finite coproducts has sum types as specified in
  Fig.~\ref{fig:polyfunctors}. It is moreover clear that $\Th(\BC)$
  has a type $\mi{Zero}=0$, operations $0\to a$ for every type $a$,
  and an axiom $\forall x:\mi{Zero}.\,\bot$, the latter because the
  unique morphism $0\to 1$ in $\BC$ equalises the truth values
  $\top,\bot:1\to\Logical$. Conversely, the classifying category of a
  theory with sums has finite coproducts: one has to check that the
  binary coproduct of objects $(x:a\obsep\phi)$, $(y:b\obsep\psi)$
  exists in the classifying category; but this is just the object
  \begin{equation*}
    (z:a+b\obsep\mi{case}\ z\ \mi{of}\ \mi{inl}\ x\to \phi\mid
    \mi{inr}\ y\to\psi).
  \end{equation*}
  Similarly, one easily checks that $0$ has a unique morphism into
  every object $(x:a\obsep\phi)$ of the classifying category. It
  follows that
  \begin{quote}
    \emph{the partial $\lambda$\dash calculus with equality and sums
      is the internal logic of pccc's with equality and finite
      coproducts}.
  \end{quote}  
  The search for the internal logic of quasitoposes, i.e.\ a logic
  that would be equivalent to quasitoposes via an internal
  language/classifying category correspondence, remains open. Recall
  that a quasitopos is a pccc with equality and finite colimits, i.e.\
  finite coproducts and coequalisers. Hence, the missing ingredient is
  a suitable logical representation of coequalisers, i.e.\
  quotients. We conjecture that the key to this is a generalisation
  from subtypes $(x:a\obsep\phi)$ to subtypes with replacement, i.e.\
  types of the form $(f(x)\obsep x:a;\phi)$, representing the quotient
  of $(x:a\obsep\phi)$ by the kernel of a function $f:a\to b$.
\end{rem}

\noindent Unlike in toposes, coproducts in quasitoposes, and hence in
pccc's with equality, need not be disjoint; i.e.\ the pullback of
distinct coproduct injections need not be the initial
object. Specifically, one has
\begin{prop}\label{prop:disj-sums}
  Let $\BC$ be a pccc with equality. Then $\BC$ has disjoint finite
  coproducts iff $\BC$ has a (strict) initial object $0$, the
  coproduct $1+1$ exists in $\BC$, and the monomorphism $0\to 1$ is
  regular.
\end{prop}
\noindent The proof needs the following observation.
\begin{lem}\label{lem:bot}
  If $0\to 1$ is regular in a pccc with equality, then every type $a$
  has a partial constant $\mi{bot}:1\pfun a$ such that
  $\neg\IsDef\ (\mi{bot}\ ())$.
\end{lem}
\begin{proof}
  The assumption implies that $0$ is isomorphic to the object
  $(x:1\obsep\bot)$. Thus we can put $\mi{bot}\ x =z\ (x\resOp\bot)$,
  where $z$ is the unique morphism $(x:1\obsep\bot)\cong 0\to a$.
\end{proof}
\begin{proof}[Proof (Proposition~\ref{prop:disj-sums})]
  \emph{`Only if'} holds universally: If coproducts are disjoint, then
  $0\to 1$ is a regular monomorphism, being the pullback e.g.\ of the
  left injection $1\to 1+1$, which is even a section. Moreover,
  initial objects in cartesian closed categories are always strict:
  since the functor $\argument\times A$ is left adjoint for every $A$,
  it preserves the initial object $0$, i.e.\ $A\times 0\cong 0$; this
  is easily seen to imply that every object $B$ that has a morphism
  $B\to 0$ is initial.
  % Namely: f:C -> 0 gives existence of morphisms; uniqueness: have
  % pi2:0->A, hence g:B->A = pi2 <f,g> . But pi1:0->0 is id, hence f =
  % pi1<f,g>=<f,g> and hence g = pi2 f.

  \emph{`If':} Given $\mi{Bool}=1+1$ with
  operations $\mi{false}$, $\mi{true}$, and
  $\mi{if}-\mi{then}-\mi{else}$ as above, one can construct the binary
  sum $a+b$ of objects $a,b$ in $\BC$ as
  \begin{equation*}
    (x:1\pfun a,y:1\pfun b,z:\mi{Bool}\obsep
    (\IsDef\ (x\ ()) \smalliff z=\mi{true}) \land 
    (\IsDef\ (y\ ())\smalliff z=\mi{false})).
  \end{equation*}
  The coproduct injections are given by $\mi{inl}\
  x=((\lambdaTerm{z:1}{x}),(\lambdaTerm{z:1}{\mi{bot}\ ()}),\mi{true})$, and
  $\mi{inr}\ y=((\lambdaTerm{z:1}{\mi{bot}\ ()}),$ $(\lambdaTerm{z:1}{y}),$
  $\mi{false})$, with $\mi{bot}$ according to Lemma~\ref{lem:bot}.
  The copairing $h=\mi{sumcase}\ f\ g$ of functions $f:a\tfun c$,
  $g:b\tfun c$ is then defined as $$h\ (x,y,z)=\mi{if}\ z\ \mi{then}\
  f\ (x\ ())\ \mi{else}\ g\ (y\ ()).$$ It is easy to see that the
  copairing is uniquely determined.

  This proves existence of finite coproducts; it remains to prove that
  coproducts are disjoint. One checks by an easy diagram chase that if
  $0$ is strict and $1+1$ is disjoint, then every coproduct is
  disjoint. It is moreover easy to see that if the monomorphism $0\to
  1$ is regular, then $1+1$ is disjoint.
  % Diagram chase: let l,r be the injections into A+B, and let C be
  % their pb with l*,r*. Then l1!l*=r1!r*, with r1, l1 injections into
  % 1+1. Thus C factors through 0, thus C=0.  
  %
  % Disjointness of 1+1: let r,l be the inj. into 1+1, and let l!=r!
  % at C; let 0->1 be the equaliser of g,f. Then (g+f)l!=(g+f)r!,
  % hence g!=f!, and hence C factors through 0 as required.
\end{proof}
\begin{rem}\label{rem:spap}
  The previous statement means in particular that coproducts in a pccc
  with equality and finite coproducts, in particular in a quasitopos,
  are disjoint iff the monomorphism $0\to 1$ is regular. This may but
  need not be the case. Positive set-based examples include the
  above-mentioned quasitoposes of pseudotopological spaces and
  (reflexive) relations, respectively, and more generally any
  set-based quasitopos whose forgetful functor preserves pullbacks and
  induces a singleton fibre over the empty set. The simplest example
  where $0\to 1$ is not regular are Heyting algebras, which are
  quasitoposes when regarded as thin categories: in these, $0$ is the
  bottom element, and $1$ is the top element, but the only regular
  monomorphisms are the isomorphisms, so that $0\to 1$ fails to be
  regular except in the degenerate case. Note that the Heyting
  algebras are, up to equivalence of categories, precisely the
  quasitoposes with an inconsistent internal logic, i.e.\ with
  $\top=\bot$. An example with a consistent internal logic is the
  following. Let $\mathrm{Spa}(\Pow)$ be the category whose objects
  are pairs $(X,\CA)$ with $X$ a set and $\CA\subseteq\Pow(X)$, and
  whose morphisms $(X,\CA)\to(Y,\CB)$ are maps $f:X\to Y$ such that
  for all $A\in\CA$, $f[A]\in\CB$. By results
  of~\cite{AdamekHerrlich90} (or easy direct verification),
  $\mathrm{Spa}(\Pow)$ is a quasitopos. However, $0\to 1$ is not
  regular: $0$ is the object $(\emptyset,\emptyset)$, while $1$ is the
  object $(\{*\},\Pow(\{*\}))$ (and hence the regular subobject
  $(z:1\obsep\bot)$ is $(\emptyset,\{\emptyset\})$, not
  $(\emptyset,\emptyset)$).
\end{rem}

\begin{rem}\label{rem:bot}
  The undefined constant $\mi{bot}$ of Lemma~\ref{lem:bot} can be used
  e.g.\ to define partial extraction functions $\mi{outl}:a+b\pfun a$,
  $\mi{outr}:a+b\pfun b$ by $\mi{outl}\ z=\mi{case}\ z\ \mi{of}\
  \mi{inl}\ x \to x\mid \mi{inr}\ y\to \mi{bot}\ ()$, analogously for
  $\mi{outr}$. (This implies moreover that $a$ and $b$ are regular
  subobjects of $a+b$.) Conversely, it should be noted that unless
  coproducts are disjoint, types $1\pfun a$ may fail to have closed
  inhabitant terms, and extraction functions need not exist. Some
  constructions in the preliminary version of this
  work~\cite{Schroder07a} erroneously made use of $\mi{bot}$ without
  identifying disjointness of coproducts as an additional
  assumption. The main results, however, remain correct also without
  this assumption, as we show below; occasionally, this requires
  slightly unexpected workarounds.
\end{rem}
\begin{rem}
  Under unique choice, $\mi{Bool}=1+1$ coincides with the type
  \begin{equation*}
    (p:\Logical\obsep p\lor\lnot p),
  \end{equation*}
  with injections $\mi{inl}\ ()=\top$ and $\mi{inr}\ ()=\bot$.  The
  copairing $h=\mi{sumcase}\ f\ g$ of two functions
  $f,g:1\tfun a$ is then defined as $h\ p=\iota x:a\sbullet
  (p\impl f\ ()=x)\land(\lnot p\impl g\ ()=x)$.
  % commutation: h (inl ()) = h T = f (), OK
  % Uniqueness: p => p = T => bar h p = bar h T =f () = h T = h p,
  % not p genauso, dann fertig per p \/ not p.
  % totality: p: x = f () erf\"ullt  Bedingung, und ist auch das einzige
  % not p: ebenso f\"ur x = g (); fertig per p \/ not p.
  Moreover, it is easy to see that, under unique choice, the object
  $(z:1\obsep\bot)$ is initial.  By Proposition~\ref{prop:disj-sums},
  this reproves the well-known fact that toposes have disjoint finite
  coproducts. In a quasitopos, one cannot in general construct
  $\mi{Bool}$ as a subtype of $\Logical$ --- the latter is typically
  an indiscrete space, while $\mi{Bool}$ is typically discrete. E.g.\
  in the quasitopos of reflexive relations, $\Logical$ carries the
  universal relation, while $\mi{Bool}$ carries the equality relation.
\end{rem}

\noindent As indicated above, we shall also need the standard notion
of natural numbers object (nno). Categorically, an nno is an initial
algebra for the functor $\argument+1$; in \HasCASL, a corresponding
type of natural numbers is specified as 
\begin{hetcasl}
\> \KW{free} \KW{type} \=\Id{Nat} \Ax{::=} \=\Ax{0} \AltBar{} \Id{suc} \Id{Nat}

\end{hetcasl}
\begin{rem}\label{rem:nno}
  In a cartesian closed category, every nno is internal in the sense
  that the unique existence of an algebra morphism from the nno into a
  given $\argument+1$\dash algebra holds as a formula of the internal logic
  and is embodied by an operation, $\mi{fold}$ in the above
  notation~\cite{Pitts00}; the same holds for initial algebras, and
  dually for final coalgebras, of arbitrary strong functors, in
  particular polynomial functors.
  % Cf. Notes.txt
  The explicit distinction of internal nno's used
  in~\cite{Schroder07a} is thus superfluous. It follows that the
  internal language of a pccc with equality, sums, and nno always has
  a type $\mi{Nat}$ as specified above. Conversely, the classifying
  category of a theory with sums and the type $\mi{Nat}$ will have
  $(n:\mi{Nat}\obsep\top)$ as an nno. To see this, one has to show
  that the fold operation applies also to $\argument+1$\dash algebras
  on subtypes $(x:a\obsep\phi)$. This can be proved without resorting
  to instances of $\mi{fold}$ at such subtypes, as indicated in
  Sect.~\ref{sec:datatypes}; as announced there, the argument
  presented in the following is general enough to apply to arbitrary
  polynomial functors. The relevance of this point is that assuming
  instances of $\mi{fold}$ at subtypes essentially amounts to
  postulating induction as a separate axiom, rather than deriving it
  from recursion.
  
  To begin, note that the induction principle on $\mi{Nat}$ may be
  proved using $\mi{fold}$ only at the type $\Logical$: given a
  predicate $P$ on $\mi{Nat}$ such that $P\ 0$ and $\forall
  n:\mi{Nat}\sbullet P\ n\impl P\ (\mi{suc}\ n)$, define a predicate
  $Q$ on $\mi{Nat}$ recursively by
  \begin{align*}
    Q\ 0 & = P\ 0\\
    Q\ (\mi{suc}\ n) & = Q\ n\land P\ (\mi{suc}\ n).
  \end{align*}
  Then $Q$ has the defining property of $\mi{fold}\ g$, where
  $g:\Logical+1\to\Logical$ is the copairing of the identity and
  $\top$. As the constantly true predicate on $\mi{Nat}$ also has this
  property, it follows that $Q$ and hence $P$ hold universally.

  Then, a morphism $d:(x:a\obsep\phi)+1\to(x:a\obsep\phi)$ induces in
  the obvious way a morphism $d^?:((1\pfun a)+1)\tfun (1\pfun a)$. One
  thus obtains $f=\mi{fold}\ d^?:\mi{Nat}\to(1\pfun a)$. It remains to
  show that $f$ factors through $(x:a\obsep\phi)$, i.e.\ that $f\ n$
  is defined and satisfies $\phi[f\ n/x]$ for all $n$; this is proved
  by induction.

  We have thus established that
  \begin{quote}
    \emph{the partial $\lambda$\dash calculus with equality, sums, and
      $\mi{Nat}$ is the internal language of pccc's with equality,
      finite coproducts, and nno}.
  \end{quote}
\end{rem}

\noindent We shall now prove the existence of initial algebras and
final coalgebras for certain classes of functors the categorical
semantics, thus partially generalising known results for $W$\dash types in
toposes (e.g.~\cite{MoerdijkPalmgren00}).
\begin{defi}
  The class of \emph{polynomial} functors is inductively generated
  from the identity functor and constant functors by taking finite
  sums and products. The class of \emph{extended polynomial functors}
  is inductively generated from the exponential functors with constant
  exponent (including the identity functor by taking exponent $1$) and
  constant functors by taking finite sums and products.
\end{defi}
\noindent Of course, the intended interpretation of the constructor
class $\mi{PolyFunctor}$ from Sect.~\ref{sec:bootstrap} is the class
of polynomial functors, and correspondingly for the more general
constructor class appearing in the semantics of cotypes
(Sect.~\ref{sec:cotypes}) and the class of extended polynomial
functors.
\begin{thm}\label{thm:main}
Let $\BC$ be a pccc with equality, finite coproducts, and
  nno (e.g.\ a quasitopos with nno). Then
\begin{enumerate}[\em(a)]
\item $\BC$ has initial algebras for polynomial functors;
\item $\BC$ has final coalgebras for extended polynomial functors. 
\end{enumerate}
\end{thm}
\noindent The constructions employed in the proof are essentially
subtype definitions in the internal logic. By the discussion in
Sect.~\ref{sec:poly} and~\ref{sec:semantics}, it follows that, as an
extension of the specification of sums and the natural numbers, the
declaration of a datatype $t=C\ a_1\ \dots\ a_n$ with constructor
arguments $t_{ij}$ as in the beginning of Sect.~\ref{sec:datatypes} is
conservative, provided that the $t_{ij}$ are built from $t$, the
$a_i$, and types from the local environment using only product and sum
type formation. Moreover, the declaration of a final process type
$t=C\ a_1\ \dots\ a_n$ as in Sect.~\ref{sec:cotypes} is conservative
as an extension of the specification of sums and the natural numbers,
provided that the codomains $t_{ij}$ of the selectors are built from
$t$, the $a_i$, and types from the local environment using only
product and sum type formation and exponentiation with exponents not
depending on $t$.  Since the class of pccc's with equality, finite
coproducts, and nno is easily seen to be stable under taking products
of categories, Theorem~\ref{thm:main} implies moreover that analogous
results hold for declarations of several mutually recursive types or
cotypes, respectively.
% nno: clear
% dominion: clear
% function spaces: clear
% coproducts: clear

We begin by proving the existence of a particular datatype, the type
of lists:
\begin{lemdefn}\label{lem:lists}
  Let $\BC$ be a pccc with equality, finite coproducts, and nno. Then
  $\BC$ has list objects, i.e.\ for every object $A$, the functor
  $1+A\times\argument$ has an initial algebra, the \emph{type of lists
    over $A$}.
\end{lemdefn}
\begin{proof}
  We construct the type $\mi{List}$ of lists over $a$ as 
  \begin{equation*}
    \mi{List}=1+(l:\mi{Nat}\pfun A,n:\mi{Nat}\obsep
    \forall m:\mi{Nat}\sbullet\IsDef\ (l\ m)\smalliff m\le n)
  \end{equation*} 
  (where $\le$ is defined recursively). We define the list constructors
  $\mi{nil}:\mi{List}$ and $\mi{cons}:A\tfun\mi{List}\tfun\mi{List}$
  by $\mi{nil}=\mi{inl}\ ()$ and 
  \begin{align*}
    \mi{cons}\ x\ l=\mi{case}\ l\
    \mi{of}\ & \mi{inl}\ () \to \mi{inr}\ ((\lambdaTerm{k:\mi{Nat}}{\mi{case}\
      k\ \mi{of}\ 0\to x\mid\mi{suc}\ m\to x\resOp\bot}),0)\\
    &\mi{inr}\ (l,n)\to \mi{inr}\ ((\lambdaTerm{k}{\mi{case}\ k\ \mi{of}\ 0\to
      x\mid \mi{suc}\ m \to l\ m}), \mi{suc}\ n).
  \end{align*}
  Given a further $1+A\times\argument$\dash algebra $B$ with
  operations $c:B$ and $f:A\times B\to B$, the folded function
  $g=\mi{fold}\ c\ f:\mi{List}\to B$ is defined by $g\ z=\mi{case}\ z\
  \mi{of}\ \mi{inl}\ ()\to c\mid\mi{inr}\ (l,n)\to h\ l\ n$,
  where $h$ is defined by recursion over $\mi{Nat}$:
  \begin{align*}
    h\ l\ 0 & = f\ (l\ 0)\ c\\
    h\ l\ (\mi{suc}\ n) & =f\ (l\ 0)\ (h\
    (\lambdaTerm{k:\mi{Nat}}{l\ (\mi{suc}\ k)})\ n).
  \end{align*}
  It is easy to check that $g$ satisfies, and is uniquely determined
  by, the defining equation for $\mi{fold}\ c\ f$. Moreover, as
  recursion on natural numbers is embodied as an operation, so is the
  recursion principle on lists.
\end{proof}

\noindent Note how in the construction of the list datatype, an
explicit list length component serves to enable inheritance of the
recursion operator from the natural numbers. This principle is also at
the heart of the general construction of datatypes below, where we
employ an explicit depth component on trees. Note moreover that this
component is \emph{not} needed in the construction of final
coalgebras. 

A maybe slightly unexpected feature of the construction, which
illustrates the points made in Remark~\ref{rem:bot}, is the fact that
we need to treat the empty list as a special case --- in general, we
cannot code it as the everywhere undefined function $\mi{Nat}\pfun A$,
as the latter may fail to exist. Singleton lists, on the other hand,
are unproblematic: once we have an element $x:A$ of the list in hand,
we obtain an undefined term of type $A$ as $x\resOp\bot$.
\begin{proof}[Proof of Theorem~\ref{thm:main}]
  By Remark~\ref{rem:nno}, we can conduct the proof in the classifying
  category of the internal language of $\BC$, the latter being a
  partial $\lambda$\dash theory with equality, finite sums, and nno.

  \emph{(a):} We can assume that the given functor $F$ is of the
  normal form $FX=\sum_{i=1}^nA_i\times X^{k_i}$ with
  $k_i\in\mathbb{N}$ and constant parameter objects $A_i$. Moreover,
  by collecting all $A_i$ with $k_i=0$ into a single sum type ($0$ in
  case $k_i>0$ for all $i$), we may assume that $k_i=0$ iff $i=1$, so
  that $A_1$ may be thought of as the type of constants in the
  signature. Let $A=\sum_{i=1}^n(A_i+1)$, with injections into the
  outer coproduct denoted $\mi{in}_i$, and injections into the inner
  coproducts denoted $\mi{inl},\mi{inr}$ as usual. Let $\mi{Path}$ be
  the type of lists of natural numbers (which exists in $\BC$
  according to Lemma~\ref{lem:lists}), with constructors
  $\mi{nil}:\mi{Path}$,
  $\mi{cons}:\mi{Nat}\tfun\mi{Path}\tfun\mi{Path}$.  We now define a
  universal type of trees, from which the desired initial algebra will
  be carved out as a subtype, by
  \begin{equation*}
    \mi{DTree}=(l:\mi{Path}\pfun A;d:\mi{Path}\pfun\mi{Nat};x:A_1),
  \end{equation*}
  where for $(l,d,x):\mi{DTree}$ and $p:\mi{Path}$, $l\ p=\mi{in_i}\
  z$ indicates that the subtree at $p$ is either a leaf labelled $y$,
  if $z=\mi{inl}\ y$, or a node labelled by the $i$\dash th
  constructor, if $z=\mi{inr}\ ()$, and $d\ p=n$ indicates that the
  subtree at $p$ has depth $n$. The third component $x:A_1$ is a dummy
  that serves only to enable the construction of undefined terms (see
  Remark~\ref{rem:bot}). We put $\mi{depth}\ (l,d,x)=d\ \mi{nil}$ for
  $(l,d,x):\mi{DTree}$, and for $j:\mi{Nat}$, $j>0$, we define a generic
  $j$\dash th selector by $\mi{sel}_j\ (l,d,x)=(l\circ (\mi{cons}\
  j),d\circ(\mi{cons}\ j),x)$. Moreover, we
  %  embed $A$ into $\mi{DTree}$ by 
  % \begin{equation*}
  %   \mi{leaf}\ x = (\lambdaTerm{p:\mi{Path}}{\mi{inl}\ x\resOp(p=\mi{nil})},
  %  \lambdaTerm{p:\mi{Path}}{0\resOp(p=\mi{nil})}), 
  % \end{equation*}
  % and 
  define generic constructors
  $c_i:A_i\times\mi{DTree}^{k_i}\to\mi{DTree}$, thus making
  $\mi{DTree}$ into an $F$-algebra, by
  \begin{equation*}
    c_i\ (y,(l_1,d_1,x_1),\dots,(l_{k_i},d_{k_i},x_{k_i})) = 
    \begin{cases}(l,d,x_1)&\textrm{if $k_i>0$}\\
      (l,d,y)& \textrm{if $k_i=0$ (and hence $i=1$, so that $y:A_1$)}
    \end{cases}
  \end{equation*}
  where $l$ and $d$ are defined by case distinction as
  \begin{align*}
    l\ \mi{nil}& =\mi{in}_i\ (\mi{inr}\ ()) & l\ (\mi{cons}\ j\ p)&=
    (\mi{if}\ j=0\ \mi{then}\ \mi{in}_i\ (\mi{inl}\ y)\ \mi{else}\ l_j\ p)\\
    d\ \mi{nil}& =1+\max\ (d_1\ \mi{nil},\dots,d_{k_i}\ \mi{nil}) & d\
    (\mi{cons}\ j\ p)&=(\mi{if}\ j=0\ \mi{then}\ 0\ \mi{else}\ d_j\ p).
  \end{align*}
  Here, the maximum is defined by recursion on the naturals, with
  $\max\ ()=0$; and the $\mi{if}$ expressions abbreviate obvious case
  expressions. The expressions denoted for the sake of readability as
  $l_j\ p$ and $d_j\ p$ on the right hand side in reality abbreviate
  long case distinctions over $j=1$, \dots, $j=k_i$, $j>k_i$, with
  $l_j\ p$ and $d_j\ p$ undefined for $j>k_i$. E.g.\ for $l_j$, we
  have
  \begin{equation*}
    l_j\ p=\begin{cases} 
      \mi{bot}_A\ y & \textrm{if $j=0$ or $j>k_i$} \\
        l_1\ p & \textrm{if $j=1$}\\
        \dots\\
        l_{k_i}\ p & \textrm{if $j=k_i$}
      \end{cases}
    \end{equation*}
  where $\mi{bot}_A\ y=(\mi{in}_i\ y)\resOp\bot$, and the case distinction
  can be emulated by a finite chain of $0/\mi{suc}$ case statements. 

  We then take the carrier $T$ of the desired initial algebra to be
  the smallest subtype of $\mi{DTree}$ closed under the $c_i$; thus,
  $T$ inherits from $\mi{DTree}$ the structure of an $F$-algebra. Note
  that for all $(l,d,x):T$, $\mi{depth}\ (l,d,x)>0$ and $l\
  \mi{nil}=\mi{in_i}\ (\mi{inr}\ ())$ for some $i$.
  % using the case operator, we may obtain the latter via
  % an operation $\mi{alt}$. 
  % We then have $l\ (\mi{cons}\ 0\ \mi{nil})=\mi{in_i}\ (\mi{inl}\ x)$
  % for some $x:A_i$, which we can access via a partial extraction
  % function $\mi{leaf}_i:\mi{DTree}\pfun A_i$.
  % the extraction function does not in general exist!
  We have to show that we can construct the function
  $\mi{fold}\ b_1\ \dots\ b_n:T\tfun B$ for functions $b_i$
  constituting an $F$-algebra on a type $B$. We define a primitive
  recursive function $f:\mi{Nat}\tfun T\pfun B$ by
  \begin{align*}
    f\ 0\ (l,d,x) & = \mi{bot}_B\ x\\
    f\ (\mi{suc}\ n)\ (l,d,x) & = 
    \begin{aligned}[t]
      &\mi{case}\ l\  \mi{nil}\ \mi{of}\\
      &\quad
        (\mi{in}_i\  z \to\ \mi{case}\ z\ \mi{of}\ 
        \mi{inl}\ y \to \mi{bot}_B\ x\\
       & \qquad \mid \mi{inr}\ () \to\ \mi{case}\ l\ [0]\ \mi{of}\\
       & \qquad\qquad
           \mi{in}_i\ y \to\mi{case}\ y\ \mi{of}\ \mi{inr}\ ()\to\mi{bot}_B\ x\\
       &  \qquad\qquad\quad \mi{inl}\ w \to b_i\ (w,f\ n\ (\mi{sel_1}\
      (l,d,x)), \dots, f\ n\ (\mi{sel}_{k_i}\ (l,d,x)))\\
       &\qquad\qquad\mi{otherwise} \to \mi{bot}_B\ x)_{i=1,\dots,n},
  \end{aligned}
  \end{align*}
  where $\mi{otherwise}$ is a placeholder for all remaining cases in a
  case statement, in this case $\mi{in}_j\ y$ for $j\neq i$,
  $\mi{bot}_B\ x=(b_1\ x)\resOp\bot$, and $[0]=\mi{cons}\ 0\
  \mi{nil}$.  Finally, put
  \begin{equation*}
    \mi{fold}\ b_1\ \dots\ b_n\ z=f\ (\mi{depth}\ z)\ z.
  \end{equation*}
  One verifies directly from the defining equations for $f$ that this
  definition satisfies the fold equation. Moreover, one shows using
  the definition of $T$ as the least subtype of $\mi{DTree}$ closed
  under the constructors that $\mi{fold}\ b_1\ \dots\ b_n$ is total
  (i.e.\ one never runs into the exceptional cases $\mi{bot}_B\ x$ in
  the above definition of $f$), and uniquely determined by the fold
  equation.
  
  \emph{(b):} We can assume that the given functor $F$ has the normal
  form $FX=\sum_{i=1}^nA_i\times (B_i\to X)$, with constant parameter
  objects $A_i$, $B_i$: it is easy to see that the class of functors
  isomorphic to such normal forms contains all exponential functors
  and all constant functors (noting that parameter objects can also be
  $1$ or $0$) and is closed under sums; to see closure under products,
  note that a product of two such normal forms is a sum of summands of
  the form $(A\times (B\to X))\times(A'\times (B'\to X))$. Such a
  summand is isomorphic to $(A\times A')\times((B+B')\to X)$.

  Now put $A=\sum_{i=1}^nA_i$ and $B=\sum_{i=1}^nB_i$, with injections
  $\mi{in}_i$ in both cases. Define $\mi{Path}$ as the type of lists
  over $B$, with constructors $\mi{nil},\mi{cons}$, and equip it with
  the standard $\mi{snoc}$ operation $\mi{Path}\times
  B\tfun\mi{Path}$. The universal type of infinite trees is
  \begin{equation*}
    \mi{PTree} = \mi{Path}\pfun A
  \end{equation*}
  (where it is crucial that we omit the depth component present in the
  universal type $\mi{DTree}$ for initial datatypes).  For
  $f:\mi{PTree}$ and $p:\mi{Path}$, the intended reading of $f\ p =
  \mi{in}_i\ x$ is that position $p$ in the tree behaves according to
  the $i$\dash th alternative and outputs $x:A_i$. The carrier of the final
  $F$\dash coalgebra is then the subtype $C$ of $\mi{PTree}$ consisting of
  those $f$ such that
  \begin{gather}
    \label{eq:Cnil} \mi{def}\ (f\ \mi{nil})\qquad\textrm{and}\\
    \label{eq:Csnoc} \mi{def}\ (f\ (\mi{snoc}\ p\ (\mi{in}_i\ y))) \smalliff 
    \exists x:A_i\sbullet f\ p =
    \mi{in}_i\ x
  \end{gather}
  for all $i=1,\dots,n$. (Note that $f\ p = \mi{in}_i\ x$ entails that
  $f\ p$ is defined.) We then define an $F$\dash coalgebra structure
  $c:C\tfun\sum_{i=1}^n(A_i\times(B_i\tfun C))$, with injections again
  denoted $\mi{in}_i$, by
  \begin{equation*}
    c\ f  =  \mi{case}\ f\ \mi{nil}\ \mi{of}\ 
    (\mi{in}_i\ x \to \mi{in}_i\ (x,
      \lambdaTotal{y:B_i}{\lambdaTerm{p:\mi{Path}}
          {f\ (\mi{cons}\ (\mi{in}_i\ y)\ p)}}))_{i=1,\dots,n}.
  \end{equation*}
  Given a further $F$\dash coalgebra $d:D\to\sum_{i=1}^n(A_i\times(B_i\tfun
  D))$, we define the morphism $u=\mi{unfold}\ d:D\to C$
  recursively by
  \begin{align*}
    u\ z\ \mi{nil} & = \mi{case}\ d\ z\ \mi{of}\ (\mi{in_i}\ (x,g)\to
    \mi{in}_i\ x)_{i=1,\dots,n}\\
    u\ z\ (\mi{cons}\ (\mi{in_i}\ y)\ p) & = 
    \mi{case}\ d\ z\ \mi{of}\ \mi{in_i}\ (x,g)\to u\ (g\ y)\ p
  \end{align*}
  where omitted cases in the second case statement are understood to
  be undefined --- we are lucky enough to have the undefined term $(u\
  z\ p)\resOp\bot$ available for this purpose (see
  Remark~\ref{rem:bot}). Since primitive recursion on lists is given
  by an operator, the above definition can be expressed as a term
  defining $\mi{unfold}\ d$ and indeed $\mi{unfold}$ as a function. It
  is immediate that $u\ z$ satisfies (\ref{eq:Cnil}); one proves by
  induction over $p$ that $u\ z$ satisfies also (\ref{eq:Csnoc}) and
  therefore indeed belongs to $C$.
  % Put is_i y = EX x. y = in_i x
  % to prove: def u z (snoc p (in_i y)) <=> is_i (u z p)
  %
  % p=nil: 
  % def u z (snoc nil (in_i y)) <=>
  % def u z (cons (in_i y) nil) <=> d_z= in_i(x,g) /\ def u (g y) nil
  % <=> is_i (d z) <=> is_i (u z nil), OK
  %
  % p=cons (in_j w) q: 
  % def u z (snoc (cons (in_j w) q) (in_i y)) <=>
  % def u z (cons (in_j w) (snoc q (in_i y))) <=>
  % def case d z of in_j (x,g) -> u (g w) (snoc (q (in_i y)))
  % <=> (IV) 
  % d z = in_j (x,g) /\ is_i u (g w) q <=> (Definition of following)
  % is_i (u z (cons (in_j w) q))
  One verifies directly that $u$ satisfies the defining equation for
  $\mi{unfold}\ d$.
  % Namely, T u (d z) = c (u z)
  % LHS: case d z of (in_i (x,g) -> in_i (x, \ y p. u (g y))
  % RHS: expand defn. of c:
  % = case u z nil of (in_i x -> (x, \ y p. u z (cons (in_i y) p))
  % expand defn of u z nil and u z (cons (in_i y) p, simplify case:
  % = case d z pf (in_i (x,g) -> (x, \ y p. u (g y) p), OK
  Finally, one shows by induction over $\mi{Path}$ that $u$ is
  uniquely determined by the unfold equation.
  % which can be decomposed (by splitting pairs and 
  % applying the right hand side to p:Path) into the conditions
  %
  % case d z of (in_i (x,g) -> in_i x) = 
  %        case u z nil of (in_i x -> in_i x) = u z nil
  % and
  % case d z of (in_i (x,g) -> \ y: B_i. u (g y) p) =
  %        case u z nil of (in_i x -> \ y: B_i. u z (cons (in_i y) p))
  %        = case d z of (in_i x -> \ y: B_i. u z (cons (in_i y) p))
  %
  % The first equation is the clause for nil in the definition of u
  % The second determines u _ (cons (in_i y) p) from u _ p.
\end{proof}

\begin{rem}\label{rem:m-types}
  The proof of Theorem~\ref{thm:main} (b) can be modified to prove
  that quasitoposes have so-called $M$\dash
  types~\cite{BergDeMarchi07}. We have omitted this aspect from the
  main line of the presentation, as it involves the use of dependent
  types (which exist in pccc's with equality~\cite{Schroder04}) and is
  not relevant for the semantics of \HasCASL. We sketch some details
  for the interested reader, who may note that, discounting the need
  for the extra machinery of dependent types, the formulation of the
  proof is in fact slightly simpler in the case of $M$\dash types.

  An $M$\dash type is defined as a final coalgebra for \emph{general
    polynomial functors}, i.e.\ functors $\Pol_q$ defined by
  \begin{equation*}
    \Pol_q(X)= \textstyle \sum a:A.\,(q^{-1}(a)\to X),
  \end{equation*}
  where $q:B\to A$ is a morphism, thought of as a dependent type
  $(B_a)_{a:A}$ with $B_a=q^{-1}(a)$, and the sum is a dependent
  sum, consisting of pairs $(a,h)$ with $a:A$ and $h:B_a\to X$; one
  has \emph{projections} $\pi_1,\pi_2$ with $\pi_1(a,h)=a$ and
  $\pi_2(a,h)=h$. 
%   (Summands $A_i\times(B_i\to X)$ as in an extended
%   polynomial functor are modelled in this context by taking $A_i$
%   copies of $B_i\to X$.) 
  One lets these $A$, $B$ play the roles of
  $A,B$, respectively, as in the proof of Theorem~\ref{thm:main}; then
  Equation~\ref{eq:Csnoc} in the definition of $C$ becomes
  \begin{equation*}
    \mi{def}\ (f\ (\mi{snoc}\ p\ b)) \smalliff 
    q\ b = f\ p.
  \end{equation*}
  The definition of the $\Pol_q$\dash coalgebra structure on $C$ is
  now
  \begin{equation*}
    c\ f  =  
    (f\ \mi{nil},
      \lambdaTotal{b:B_{f\,\mi{nil}}}{\lambdaTerm{p:\mi{Path}}
          {f\ (\mi{cons}\ b\ p)}}).
  \end{equation*}
  Finally, the unique coalgebra morphism from $u=\mi{unfold}\ d:D\to
  C$ from a further $\Pol_q$\dash coalgebra $d:D\to\Pol_q(D)$ into $(C,c)$
  is recursively defined by
  \begin{align*}
        u\ z\ \mi{nil} & = \pi_1\ (d\ z)\\
    u\ z\ (\mi{cons}\ b\ p) & = 
    u\ (\pi_2\ (d\ z)\ b)\ p
  \end{align*}
  on the understanding that $B_a$ is a subtype of $B$ and that
  functions $B_a\to X$ (such as $\pi_2\ (d\ z):B_{\pi_1\,(d\,z)}\to
  X$) extend to partial functions $B\pfun X$ defined precisely on
  $B_{a}$ (this is in agreement with the coding of dependent types in
  pccc's with equality according to~\cite{Schroder04}).  The proof
  thus modified establishes that \emph{pccc's with equality, finite
    coproducts, and nno, in particular quasitoposes with nno, have
    $M$-types}.

  The diligent reader may wonder where the minor trick went that we
  had to apply in the construction of $u$ in the proof of
  Theorem~\ref{thm:main} (b): recall that we needed an undefined term
  of type $A$, which we obtained as $(u\ z\ p)\resOp\bot$. No such
  thing is needed above (although, of course, the term is still
  available). The answer to this puzzle is that unless coproducts are
  disjoint, which by Lemma~\ref{lem:bot} gives us a constant
  $\mi{bot}:1\pfun a$ at every type $a$, $M$\dash types do not
  generalise the final coalgebras of Theorem~\ref{thm:main}, since
  general polynomial functors do not actually generalise extended
  polynomial functors in the sense of the theorem. As a simple
  example, let $A=1+1$ and $B=B_l+B_r$, with the obvious projection
  $q:B\to A$. Theorem~\ref{thm:main} yields a final coalgebra for the
  extended polynomial functor $F(X)=(B_l\to X)+(B_r\to X)$, while the
  $M$\dash type considered above is a final coalgebra for the functor
  $\Pol_q(X)=\sum a:A. (q^{-1}(a)\to X)$. Although one tends to
  believe that the two functors should be isomorphic, this is not in
  general true unless coproducts are disjoint. To see this, consider
  the construction of $\sum a:A. (q^{-1}(a)\to X)$ according
  to~\cite{Schroder04}, which is just a reformulation of the natural
  set-theoretic description:
  \begin{equation*}
    \Pol_q(X)=(f:B\pfun X;a:A\obsep 
    \forall b:B.\,\IsDef\ (f\ b)\smalliff q\ b=a).
  \end{equation*}
  Using $\mi{bot}$, we can construct an isomorphism
  $h:F(X)\to\Pol_q(X)$ by
  \begin{equation*}
    h\ (\mi{inl}\ f) = ((\lambdaTerm{b:B}{\mi{case}\ b\ \mi{of}\ 
      \mi{inl}\ x \to f\ x\mid
      \mi{inr}\ y \to\mi{bot}\ ()}),\mi{inl}\ ()),
  \end{equation*}
  analogously on the other summand, but without $\mi{bot}$, the
  construction of $h$ is not possible. Indeed, a simple example shows
  that $\Pol_q$ and $F$ need not be isomorphic in general. Recall the
  category $\mathrm{Spa}(\Pow)$ from Remark~\ref{rem:spap}, which has
  non-disjoint coproducts. Let $B_l$ be the object
  $1=(\{*\},\Pow(\{*\}))$ of this category, and let $B_r$ be the
  object $1_\emptyset=(\{*\},\emptyset)$. Then $F$ as above has
  $F(1_\emptyset)=(1\to 1_\emptyset)+(1_\emptyset\to
  1_\emptyset)\neq\emptyset$ (the right hand summand contains the
  identity map). However, $\Pol_q(1_\emptyset)=\emptyset$, as there is
  no partial morphism $f:B=B_l+B_r=1+1_\emptyset\pfun 1_\emptyset$,
  because the structure of $1+1_\emptyset$ contains the empty subset.

  Both part (b) of Theorem~\ref{thm:main} and the above existence
  proof for $M$\dash types in quasitoposes complement recent results
  of van den Berg and De Marchi~\cite{BergDeMarchi07} (extending
  earlier work by Santocanale~\cite{Santocanale03}), which live in the
  setting of locally cartesian closed categories with disjoint
  coproducts and nno. In particular, existence of $M$\dash types in
  quasitoposes \emph{with disjoint coproducts}, while not formally
  stated in~\cite{BergDeMarchi07}, would seem to follow by a
  straightforward adaptation of the arguments used there
  (specifically, exchange decidable subobjects for regular subobjects
  in the proof of Proposition~4.4 in~\cite{BergDeMarchi07}). We point
  out that in our setting, we obtain a comparatively simple
  construction of $M$-types --- in the presence of partial function
  types, one can write down the $M$-type directly as a type of certain
  partial functions on paths, while the framework
  of~\cite{BergDeMarchi07} requires a more roundabout approach
  involving in particular the construction of infinite trees as
  sequences of finite-depth approximations.
\end{rem}

\begin{rem}
  The crucial difference between the above proof and the constructions
  of~\cite{Paulson97,BerghoferWenzel99}, which are also the basis of
  the topos-theoretic arguments in~\cite{MoerdijkPalmgren00}, is the
  definition of the universal types as partial function spaces rather
  than types of sets of nodes, reflecting the fact that functional
  relations need not be functions in the absence of unique
  choice. Moreover, the construction of primitive recursive functions
  can no longer rely on an inductive construction of their graphs. It
  is an open problem whether our use of the depth function for this
  purpose in the case of initial datatypes can be generalised so as to
  cover also infinitely branching datatypes such as the type
  $\mi{Tree}\ a\ b$ from Example~\ref{expl:freetypes}, or more
  generally \emph{$W$-types}, i.e.\ initial algebras for general
  polynomial functors $\Pol_q$ as in Remark~\ref{rem:m-types} (such
  types do exist in toposes with
  nno~\cite{PareSchumacher78,MoerdijkPalmgren00}; even more generally,
  the existence of $W$-types implies the existence of initial algebras
  for dependent polynomial functors~\cite{GambinoHyland03}).
\end{rem}

\section{Domains}\label{sec:domains}

\noindent The treatment of general recursion in \HasCASL is based on a
HOLCF-style~\cite{Regensburger95} internal representation of domains,
phrased in terms of chain-complete partial orders. Some adaptations to
this theory are necessary in order to cope with the absence of unique
choice~\cite{SchroderMossakowski08HCOverview}. We briefly recall the
relevant definitions and results below, and then go on to discuss the
existence of initial datatypes in the category of domains. As already
in the case of datatypes, we work in the internal language of a pccc
with equality, sums, and nno; \emph{additionally, we assume
  disjointness of coproducts}.

The main difficulty is that without unique choice, we can no longer
e.g.\ define the value at $x$ of the supremum of a chain of partial
functions $f_i$ as `the value (if any) eventually assumed by the
$f_i(x)$'. Hence the modified definition
\begin{defi} 
  A partial order $A$ with ordering $\infle$ is called a
  \emph{complete partial order (cpo)} if the type $1\pfun A$, equipped
  with the ordering
\begin{equation*}
x\infle y \smalliff (\DEF\  x\ ()\impl x\ ()\infle y\ ()),
\end{equation*}
has suprema of chains, denoted by $\cposup$, and a bottom element
$\bot$ (the latter is not, of course, a bottom element of $A$
itself). We call chains in $1\pfun A$ \emph{partial chains}. We say
that a cpo $A$ is \emph{pointed} (or a \emph{cppo}) if $A$ has a
bottom element. We say that $A$ is a \emph{flat cpo} if $A$ is a cpo
when equipped with the discrete ordering. A partial function between
cpo's is \emph{continuous} iff it preserves suprema of partial
chains. The types of total and partial continuous functions from $A$
to $B$ are denoted $A\cArrow B$ and $A\pcArrow B$, respectively.
\end{defi}
\begin{lem}\label{lem:partial-chains}
Let~$(x_i)$ be a partial chain. Then $\cposup_i x_i$ is defined
iff $\exists n\sbullet \DEF\ x_n$.
\end{lem}
\noindent Cpo's can be specified as a class in \HasCASL; this is
carried out in detail in~\cite{SchroderMossakowski08HCOverview}. While
under unique choice, all types can be made into flat cpo's, this need
not be the case without unique choice.  Cppo's in the above sense have
least fixed points of continuous endofunctions $f$, constructed as
suprema of (total) chains $(f^n\bot)$; this is the basis of the
interpretation of general recursive functions. 
Cpo's are closed under
the usual type constructors:

\begin{prop}
Let $A$ and $B$ be cpo's. Then $A\times B$, equipped with the
componentwise ordering, is a cpo.
\end{prop}

\begin{prop}
  Let $A$ and $B$ be cpo's, and let $C$ be a type. Then the types
  $C\tfun B$, $C\pfun B$, $A\cArrow B$, and $A\pcArrow B$ are cpo's
  when equipped with the componentwise ordering; $C\pfun B$ and
  $A\pcArrow B$ are moreover pointed.
\end{prop}
\begin{prop}
The unit type is a cpo.
\end{prop}
\begin{cor}
  If $A$ is a cpo, then $1\pfun A$ is a cppo.
\end{cor}

\noindent In general, the sum of two cpo's, even $\mi{Bool}=1 + 1$,
need not be a cpo when equipped with the sum ordering. However, we
have
\begin{lem}\label{lem:cpo-sums}
  Cpo's are stable under sums of partial orders iff $\mi{Bool}$ is a
  flat cpo.
\end{lem}
\begin{rem}
  The previous lemma is the crucial point where disjointness of
  coproducts (in the shape of $\mi{bot}$) is needed. One could
  alternatively just assume that cpo's are stable under sums of
  partial orders, but this is conceptually not entirely satisfactory.
\end{rem}
\noindent The syntactic sugaring of domains in \HasCASL includes a
{\bf free domain} construct that declares initial algebras in the
category of cpo's and continuous functions (rather than in the
category of types and functions as in the case of {\bf free type}).
We now show that the initial datatypes and final process types for
polynomial and extended polynomial functors $F$, respectively,
constructed in the proof of Theorem~\ref{thm:main} can be made
(respectively, in the case of final process types, slightly modified)
into cpo's in such a way that they become initial algebras and final
coalgebras, respectively, for the corresponding functor, denoted $\bar
F$, on the category of cpo's and continuous functions, where in the
case of extended polynomial functors, function spaces are replaced by
continuous function spaces. It is an important open problem whether
this result can be extended to datatypes $t$ with non-strict
constructors, i.e.\ with arguments of type~$1\pfun t$, such as the
type of lazy lists.  In the following, we assume that \emph{$\mi{Nat}$
  is a flat cpo} (this may or may not be the case in concrete
models~\cite{SchroderMossakowski08HCOverview}); consequently,
$\mi{Bool}$ is also a flat cpo, and hence cpo's are stable under sums
by Lemma~\ref{lem:cpo-sums}.

\subsection*{Initial Datatypes as Cpo's}
Let $T$ be the initial algebra for the functor
$FX=\sum_{i=1}^n(A_i\times X^{k_i})$ as in Sect.~\ref{sec:cons}, where
the parameter objects $A_i$ are cpo's. Then the ordering on $T$ is
inherited, reusing here and below the notation from the proof of
Theorem~\ref{thm:main}, from $\mi{DTree}$ (this is equivalent to the
obvious recursive definition of a componentwise ordering), which by
the above results and under the given assumptions is a cppo.
\begin{prop}
  With the above ordering, $T$ is an initial $\bar F$\dash algebra in the
  category of cpo's and continuous functions.
\end{prop}
\begin{proof}
  It is easy to see that the constructors $c_i$ as defined in the
  proof of Theorem~\ref{thm:main} are continuous. To prove that $T$ is
  a cpo, it suffices to show that the supremum in $\mi{DTree}$ of a
  partial chain $s$ in $T$ is again in $T$, provided that $\sup s$ is
  a defined value in $\mi{DTree}$. We proceed by induction over
  $\mi{depth}\ \cposup s_m$. Let $s_m=(l_m,d_m)$ for all $m$, and let
  $\sup s_m=(l,d)$. Then $l\ \mi{nil}=\mi{in}_i\ ()$ for some $i$. By
  the definition of the sum ordering and
  Lemma~\ref{lem:partial-chains}, there is some $m$ such that $l_r\
  \mi{nil}=\mi{in_i}\ ()$ for all $r\geq m$. Since $l_r$ is in $T$, we
  have $s_r=c_i\ (\mi{leaf}_i\ s_r,\mi{sel}_1\
  s_r,\dots,\mi{sel}_{k_i}\ s_r)$ and $\mi{sel}_j\ s_r:T$ for
  $j=1,\dots,k_i$ and $r\geq m$. By continuity of $c_i$, it now follows
  from the inductive assumption that $\sup s_m$ belongs to $T$.

  It remains to be shown that for continuous functions $b_i$
  representing a $\bar F$\dash algebra on a cpo $B$, the function
  $\mi{fold}\ b_1\ \dots\ b_n:T\tfun B$ is continuous. It is easy to
  see that, given the auxiliary function $f:\mi{Nat}\tfun T\pfun B$
  from the proof of Theorem~\ref{thm:main}, the function $f\ n$ is
  continuous for every $n$ in $\mi{Nat}$. Since $\mi{Nat}$ is equipped
  with the flat ordering, it follows that $f$ itself is
  continuous. Continuity of $\mi{fold}\ b_1\ \dots\
  b_n=\lambdaTerm{z}{f\ (\mi{depth}\ z)\ z}$ then follows by the
  (obvious) continuity of $\mi{depth}$. In fact, $f$ even depends
  continuously on the $b_i$, so that $\mi{fold}$ itself is continuous.
\end{proof}
\noindent We have thus established that
\begin{quote}
  \emph{the category of cpo's in a pccc with equality, disjoint finite
    coproducts, and nno has initial algebras of polynomial functors if
    the nno is a flat cpo},
\end{quote}
and hence that declarations of {\bf free domains} for polynomial
functors in \HasCASL are conservative as extensions of the
specification of sums and a flat cpo of natural numbers; moreover, the
above proof shows additionally that the fold operator, and hence the
primitive recursion operator, is a continuous higher order function.

\subsection*{Final Process Types as Cpo's}

Unlike in the case of initial datatypes, we have to modify the
universal type $\mi{PTree}$ to the type
\begin{equation*}
  \mi{CPTree}=\mi{Path}\pcArrow A,
\end{equation*}
again reusing the notation from the proof of Theorem~\ref{thm:main},
in order to obtain a coalgebra structure for $\bar F$. By the above
results, including the fact that list types are cpo's, $\mi{CPTree}$ is
a cppo. The definition of the subtype $C$, the structure map
$c:C\tfun\bar F C$, and the function $u=\mi{unfold}\ d:D\tfun C$ for a
continuous $\bar F$\dash coalgebra $d$ on a cpo $D$ are otherwise literally
the same as in the proof of Theorem~\ref{thm:main}. It is easy to see
that $C$ is closed under suprema of chains in $\mi{CPTree}$ and hence
a cpo. Since $C$ consists of continuous maps, $c\ f$ is really in
$\bar F C$ (where functions $B_i\cArrow C$ must be continuous) for
$f:C$. It is straightforward to check that $c$ and $u$ are continuous,
using in the latter case the fact established above that primitive
recursive functions (here, on $\mi{Path}$), as well as the primitive
recursion operator itself, are continuous. We have thus shown that
\begin{quote}
  \emph{the category of cpo's in a pccc with equality, disjoint finite
    coproducts, and nno has final coalgebras of extended polynomial
    functors if the nno is a flat cpo}
\end{quote}
and hence that corresponding declarations of final process types as
cpo's for extended polynomial functors in \HasCASL are conservative as
extensions of the specification of sums and a flat cpo of natural
numbers. (Recall that such declarations are not a \HasCASL language
feature as such, but can be emulated according to
Sect.~\ref{sec:bootstrap} and~\ref{sec:cotypes}.)

\section{Conclusion}

\noindent We have laid out how initial datatypes and final process
types are incorporated into \HasCASL, and we have established the
existence of such types for a broad class of signature formats. The
main contribution in the latter respect is the avoidance of the unique
choice principle, which means that, on a more abstract level, our
constructions work in any quasitopos (more precisely, in any partial
cartesian closed category with equality and finite coproducts) with a
natural numbers object. We have moreover discussed how the
constructions can be adapted to yield corresponding types with a
domain structure as used in \HasCASL's internal modelling of general
recursion.

We have remarked that our construction of final process types can be
modified to prove existence of so-called
$M$-types~\cite{BergDeMarchi07}, i.e.\ final coalgebras for general
polynomial functors, defined over signatures given in terms of an
arbitrary morphisms $q:B\to A$. While toposes with nno also have
$W$-types, i.e.\ initial algebras for such
functors~\cite{MoerdijkPalmgren00}, the extension of our construction
of initial datatypes beyond finite branching remains an open
problem. A further point of interest for future research are datatypes
with lazy constructors, such as the type of lazy lists, in
quasitoposes, and in particular in the category of internal cpo's in a
quasitopos.  Support for datatypes with finitary polynomial signatures
is already implemented in the heterogeneous tool set
Hets~\cite{MossakowskiEA07}; support for more complex signatures,
intertwined with \HasCASL's type class mechanism as described here, is
forthcoming.

\section*{Acknowledgements}
\noindent The author wishes to thank Till Mossakowski, Christoph Lüth,
Christian Maeder, and Bernd Krieg-Br\"uckner for collaboration on
\HasCASL, Peter Johnstone and Sam Staton for helpful remarks, and the
anonymous referees for their valuable suggestions for improvement of
the paper. In particular, the second referee has provided useful hints
on $M$-types and disjointness of coproducts. Moreover, Erwin R.\
Catesbeiana has voiced his opinion on conservative extensions.

\bibliographystyle{myabbrv} \bibliography{hascasltypes}

\begin{thebibliography}{10}

\bibitem{AdamekHerrlich90}
J.~Ad{\'a}mek, H.~Herrlich, and G.~E. Strecker.
\newblock {\em Abstract and Concrete Categories}.
\newblock Wiley Interscience, 1990.

\bibitem{BerghoferWenzel99}
S.~Berghofer and M.~Wenzel.
\newblock Inductive datatypes in {HOL} - lessons learned in formal-logic
  engineering.
\newblock In Y.~Bertot, G.~Dowek, A.~Hirschowitz, C.~Paulin, and L.~Th{\'e}ry,
  eds., {\em Theorem Proving in Higher Order Logics, TPHOLs 1999}, vol. 1690 of
  {\em Lect.\ Notes Comput.\ Sci.}, pp. 19--36. Springer, 1999.

\bibitem{CASL-UM}
M.~Bidoit and P.~D. Mosses.
\newblock {\em \textsc{Casl} User Manual}, vol. 2900 of {\em Lect.\ Notes
  Comput.\ Sci.}
\newblock Springer, 2004.

\bibitem{BirkedalMoegelberg05}
L.~Birkedal and R.~E. M{\o}gelberg.
\newblock Categorical models for {Abadi} and {Plotkin's} logic for
  parametricity.
\newblock {\em Math.\ Struct.\ Comput.\ Sci.}, 15, 2005.

\bibitem{GambinoHyland03}
N.~Gambino and M.~Hyland.
\newblock Wellfounded trees and dependent polynomial functors.
\newblock In S.~Berardi, M.~Coppo, and F.~Damiani, eds., {\em Types for Proofs
  and Programs, TYPES 2003}, vol. 3085 of {\em Lect.\ Notes Comput.\ Sci.}, pp.
  210--225. Springer, 2003.

\bibitem{Girard1989}
J.-Y. Girard.
\newblock {\em Proofs and Types}.
\newblock Cambridge University Press, 1989.
\newblock Translated and with appendices by {P}. {T}aylor and {Y}. {L}afont.

\bibitem{Gunter93}
E.~L. Gunter.
\newblock A broader class of trees for recursive type definitions for {HOL}.
\newblock In J.~J. Joyce and C.-J.~H. Seger, eds., {\em Higher Order Logic
  Theorem Proving and Its Applications, HUG 1993}, vol. 780 of {\em Lect.\
  Notes Comput.\ Sci.}, pp. 141--154. Springer, 1993.

\bibitem{KiselyovEA04}
O.~Kiselyov, R.~L{\"a}mmel, and K.~Schupke.
\newblock Strongly typed heterogeneous collections.
\newblock In H.~Nilsson, ed., {\em Haskell Workshop, Haskell 2004}, pp.
  96--107. ACM, 2004.

\bibitem{LambekScott86}
J.~Lambek and P.~J. Scott.
\newblock {\em Introduction to Higher-Order Categorical Logic}.
\newblock Cambridge, 1986.

\bibitem{MoerdijkPalmgren00}
I.~Moerdijk and E.~Palmgren.
\newblock Wellfounded trees in categories.
\newblock {\em Ann.\ Pure Appl.\ Logic}, 104:189--218, 2000.

\bibitem{Moggi86}
E.~Moggi.
\newblock Categories of partial morphisms and the {$\lambda_p$}-calculus.
\newblock In D.~H. Pitt, S.~Abramsky, A.~Poign{\'e}, and D.~E. Rydeheard, eds.,
  {\em Category Theory and Computer Programming}, vol. 240 of {\em Lect.\ Notes
  Comput.\ Sci.}, pp. 242--251. Springer, 1986.

\bibitem{MossakowskiEA07}
T.~Mossakowski, C.~Maeder, and K.~L{\"u}ttich.
\newblock The {H}eterogeneous {T}ool {S}et.
\newblock In O.~Grumberg and M.~Huth, eds., {\em Tools and Algorithms for the
  Construction and Analysis of Systems, TACAS 2007}, vol. 4424 of {\em Lect.\
  Notes Comput.\ Sci.}, pp. 519--522. Springer, 2007.

\bibitem{MossakowskiEA04}
T.~Mossakowski, L.~Schr{\"o}der, M.~Roggenbach, and H.~Reichel.
\newblock Algebraic-co-algebraic specification in {\CoCASL}.
\newblock {\em J.\ Logic Algebraic Programming}, 67:146--197, 2006.

\bibitem{CASL-RM}
P.~D. Mosses, ed.
\newblock {\em {\CASL} Reference Manual}, vol. 2960 of {\em Lect.\ Notes
  Comput.\ Sci.}
\newblock Springer, 2004.

\bibitem{NipkowEA02}
T.~Nipkow, L.~C. Paulson, and M.~Wenzel.
\newblock {\em Isabelle/{HOL} - {A} Proof Assistant for Higher-Order Logic},
  vol. 2283 of {\em Lect.\ Notes Comput.\ Sci.}
\newblock Springer, 2002.

\bibitem{PareSchumacher78}
R.~Par{\'e} and D.~Schumacher.
\newblock Abstract families and the adjoint functor theorems.
\newblock In P.~Johnstone and R.~Par{\'e}, eds., {\em Indexed categories and
  their applications}, vol. 661 of {\em Lect.\ Notes Math.}, pp. 1--125.
  Springer, 1978.

\bibitem{Paulson95}
L.~C. Paulson.
\newblock Set theory for verification. {II}: Induction and recursion.
\newblock {\em J.\ Autom.\ Reasoning}, 15:167--215, 1995.

\bibitem{Paulson97}
L.~C. Paulson.
\newblock Mechanizing coinduction and corecursion in higher-order logic.
\newblock {\em J.\ Log.\ Comput}, 7:175--204, 1997.

\bibitem{Phoa92}
W.~Phoa.
\newblock An introduction to fibrations, topos theory, the effective topos and
  modest sets.
\newblock Research report ECS-LFCS-92-208, Lab.\ for Foundations of Computer
  Science, University of Edinburgh, 1992.

\bibitem{Pitts00}
A.~Pitts.
\newblock Categorical logic.
\newblock In S.~Abramsky, D.~Gabbay, and T.~Maibaum, eds., {\em Handbook of
  Logic in Computer Science}, vol. 5, Algebraic and Logical Structures,
  chapter~2. Oxford University Press, 2000.

\bibitem{Regensburger95}
F.~Regensburger.
\newblock {HOLCF}: Higher order logic of computable functions.
\newblock In E.~T. Schubert, P.~J. Windley, and J.~Alves-Foss, eds., {\em
  Theorem Proving in Higher Order Logics, TPHOLS 1995}, vol. 971 of {\em Lect.\
  Notes Comput.\ Sci.}, pp. 293--307, 1995.

\bibitem{RosoliniThesis}
G.~Rosolini.
\newblock {\em Continuity and Effectiveness in Topoi}.
\newblock PhD thesis, Merton College, Oxford, 1986.

\bibitem{RosoliniStreicher99}
G.~Rosolini and T.~Streicher.
\newblock Comparing models of higher type computation.
\newblock In L.~Birkdedal, J.~van Oosten, G.~Rosolini, and D.~S. Scott, eds.,
  {\em Realizability Semantics and Applications}, vol.~23 of {\em Electron.\
  Notes Theoret.\ Comput.\ Sci.}, 1999.

\bibitem{Santocanale03}
L.~Santocanale.
\newblock Logical construction of final coalgebras.
\newblock In H.~P. Gumm, ed., {\em Coalgebraic Methods in Computer Science,
  CMCS 2003}, vol.~82 of {\em Electron.\ Notes Theoret.\ Comput.\ Sci.}
  Elsevier, 2003.

\bibitem{Schroder04}
L.~Schr{\"o}der.
\newblock The logic of the partial {$\lambda$}-calculus with equality.
\newblock In J.~Marcinkowski and A.~Tarlecki, eds., {\em Computer Science
  Logic}, vol. 3210 of {\em Lect.\ Notes Comput.\ Sci.}, pp. 385--399.
  Springer, 2004.

\bibitem{Schroder06a}
L.~Schr{\"o}der.
\newblock The {\HasCASL} prologue - categorical syntax and semantics of the
  partial {$\lambda$}-calculus.
\newblock {\em Theoret.\ Comput.\ Sci.}, 353:1--25, 2006.

\bibitem{Schroder07a}
L.~Schr{\"o}der.
\newblock Bootstrapping types and cotypes in {\HasCASL}.
\newblock In T.~Mossakowski and U.~Montanari, eds., {\em Algebra and Coalgebra
  in Computer Science, CALCO 2007}, vol. 4624 of {\em Lect.\ Notes Comput.\
  Sci.}, pp. 447--462. Springer, 2007.

\bibitem{SchroderMossakowski08HCOverview}
L.~Schr{\"o}der and T.~Mossakowski.
\newblock {\HasCASL}: Integrated higher-order specification and program
  development.
\newblock {\em Theoret.\ Comput.\ Sci.}
\newblock To appear.

\bibitem{SchroderMossakowski02}
L.~{Schr\"oder} and T.~Mossakowski.
\newblock {\HasCASL}: Towards integrated specification and development of
  {Haskell} programs.
\newblock In H.~Kirchner and C.~Ringeissen, eds., {\em Algebraic Methodology
  and Software Technology, AMAST 2002}, vol. 2422 of {\em Lect.\ Notes Comput.\
  Sci.}, pp. 99--116. Springer, 2002.

\bibitem{SchroderMossakowski03a}
L.~Schr{\"o}der and T.~Mossakowski.
\newblock Monad-independent {Hoare} logic in {\HasCASL}.
\newblock In M.~Pezz{\`e}, ed., {\em Fundamental Approaches to Software
  Engineering, FASE 2003}, vol. 2621 of {\em Lect.\ Notes Comput.\ Sci.}, pp.
  261--277. Springer, 2003.

\bibitem{SchroderMossakowski04b}
L.~Schr{\"o}der and T.~Mossakowski.
\newblock Generic exception handling and the {J}ava monad.
\newblock In C.~Rattray, S.~Maharaj, and C.~Shankland, eds., {\em Algebraic
  Methodology and Software Technology, AMAST 2004}, vol. 3116 of {\em Lect.\
  Notes Comput.\ Sci.}, pp. 443--459. Springer, 2004.

\bibitem{SchroderMossakowski04}
L.~Schr{\"o}der and T.~Mossakowski.
\newblock Monad-independent dynamic logic in {\HasCASL}.
\newblock {\em J.\ Logic Comput.}, 14:571--619, 2004.

\bibitem{SchroderEA04}
L.~Schr{\"o}der, T.~Mossakowski, and C.~L{\"u}th.
\newblock Type class polymorphism in an institutional framework.
\newblock In J.~Fiadeiro, ed., {\em Recent Developments in Algebraic
  Development Techniques, 17th International Workshop, WADT 04}, vol. 3423 of
  {\em Lect.\ Notes Comput.\ Sci.}, pp. 234--248. Springer, 2004.

\bibitem{BergDeMarchi07}
B.~van~den Berg and F.~{De Marchi}.
\newblock Non-well-founded trees in categories.
\newblock {\em Ann.\ Pure Appl.\ Logic}, 146:40--59, 2007.

\bibitem{WalterEA05}
D.~Walter, L.~Schr{\"o}der, and T.~Mossakowski.
\newblock Parametrized exceptions.
\newblock In J.~Fiadeiro and J.~Rutten, eds., {\em Algebra and Coalgebra in
  Computer Science, CALCO 05}, vol. 3629 of {\em Lect.\ Notes Comput.\ Sci.},
  pp. 424--438. Springer, 2005.

\bibitem{Wyler91}
O.~Wyler.
\newblock {\em Lecture notes on topoi and quasitopoi}.
\newblock World Scientific, 1991.

\end{thebibliography}

\end{document}